%% file: online_learning_v8_double_column_NoHighlight.tex
\documentclass[journal]{IEEEtran}
\usepackage{cite}

\ifCLASSINFOpdf
\else
\fi
\usepackage{amsmath}
\usepackage{amsfonts}
\usepackage{algorithm}
\usepackage{algorithmic}

\ifCLASSOPTIONcompsoc
  \usepackage[caption=false,font=normalsize,labelfont=sf,textfont=sf]{subfig}
\else
  \usepackage[caption=false,font=footnotesize]{subfig}
\fi

\usepackage{graphicx}
\usepackage{url}

\usepackage{multirow}
\usepackage{array}

\usepackage{colortbl}
\usepackage[dvipsnames]{xcolor}
\usepackage{hhline}

\include{new_command}

\newcommand*{\Btr}{B_\mathrm{tr}}
\newcommand*{\Grisk}{\Gamma_\mathrm{risk}}
\newtheorem{theorem}{Theorem}

\newtheorem{lemma}{Lemma}

\newcommand{\redtext}[1]{{\color{black} #1}}

\usepackage{eqparbox}

\usepackage{etoolbox}  
\makeatletter
\patchcmd{\algorithmic}{\addtolength{\ALC@tlm}{\leftmargin} }{\addtolength{\ALC@tlm}{\leftmargin}}{}{}
\makeatother

\begin{document}
%
\title{Online Learning for Position-Aided Millimeter Wave Beam Training}
%
%
%

\author{ Vutha Va,~\IEEEmembership{Student Member,~IEEE,} Takayuki Shimizu,~\IEEEmembership{Member,~IEEE,} Gaurav Bansal,~\IEEEmembership{Member,~IEEE,} and~Robert W. Heath, Jr.,~\IEEEmembership{Fellow,~IEEE,}
\thanks{Vutha Va and Robert W. Heath, Jr. are with the Wireless Networking and Communications Group, The University of Texas at Austin, TX 78712-1687 USA (e-mail: vutha.va@utexas.edu, rheath@utexas.edu).}
\thanks{Takayuki Shimizu and Gaurav Bansal are with TOYOTA InfoTechnology Center, U.S.A., Inc., Mountain View, CA 94043 USA (e-mail: tshimizu@us.toyota-itc.com, gbansal@us.toyota-itc.com).}
\thanks{V. Va and R. W. Heath, Jr. were supported in part by the U.S. Department of Transportation through the Data-Supported Transportation Operations and Planning (D-STOP) Tier 1 University Transportation Center and by the Texas Department of Transportation under Project 0-6877 entitled ``Communications and Radar-Supported Transportation Operations and Planning (CAR-STOP)" and by a gift from TOYOTA InfoTechnology Center, U.S.A., Inc.}
}

%
%

\markboth{For submission to IEEE ACCESS: 
}%
{Va \MakeLowercase{\textit{et al.}}: Online Learning for Position-Aided Millimeter Wave Beam Training}
%



\maketitle

\begin{abstract}
Accurate beam alignment is essential for beam-based millimeter wave communications. 
Conventional beam sweeping solutions often have large overhead, which is unacceptable for mobile applications like vehicle-to-everything.
Learning-based solutions that leverage sensor data like position to identify good beam directions are one approach to reduce the overhead. 
Most existing solutions, though, are supervised-learning where the training data is collected beforehand. In this paper, we use a multi-armed bandit framework to develop online learning algorithms for beam pair selection and refinement. The beam pair selection algorithm learns coarse beam directions in some predefined beam codebook, e.g., in discrete angles separated by the 3dB beamwidths. The beam refinement fine-tunes the identified directions to match the peak of the power angular spectrum at that position. The beam pair selection uses the upper confidence bound (UCB) with a newly proposed risk-aware feature, while the beam refinement uses a modified optimistic optimization algorithm. The proposed algorithms learn to recommend good beam pairs quickly. When using 16x16 arrays at both the transmitter and receiver, it can achieve on average 1dB gain over the exhaustive search (over 271x271 beam pairs) on the unrefined codebook within 100 time-steps with a training budget of only 30 beam pairs. 
\end{abstract}


%
\IEEEpeerreviewmaketitle

\section{Introduction}
%
%
%
%


Position information may be leveraged for fast beam alignment in millimeter wave (mmWave) systems \cite{Inverse-fingerprint-preprint,Devoti2016,Aviles2016}. Such side information is widely available in vehicular applications of mmWave \cite{mmwave-vehicular-survey}. Inverse fingerprinting is one approach to exploit position information \cite{Inverse-fingerprint-preprint}, which works in non-line-of-sight (NLOS) channels. The key idea is that machine learning (ML) is used to make recommendations of promising beam pairs based on the location of the target vehicle relative to the base station (BS) and past beam measurements. \redtext{ 
The intuition here is drawn from the recommender system analogy where the user ID corresponds to the relative location and the past user ratings correspond to past beam measurement results. Given the location, past beam measurements can be input into a learning algorithm that learns to rank promising beam directions. By prioritizing beam training in top-ranked directions, the training overhead can be reduced. 
In other words, the location information is used to condition the environment, and past beam measurements in the location are used to learn which directions are promising in that environment. 
}

Although efficient, the inverse fingerprinting method has some limitations. First, the approach is offline, which means its use is delayed until the database is collected. Second, also due to being offline, its performance depends entirely on the accuracy of the collected database, which may become stale over time. Online approaches keep collecting new observations during operation, making it possible to improve the database.   
Third, without any knowledge of the power angular spectrum (PAS), the codebook must uniformly cover the antenna array's field of view (e.g., beams are spaced by the 3dB beamwidths). At a given location, depending on the scatterers in the environment, the PAS will have peaks at some specific angles. By adapting the beams such that their main beam directions match those peaks, we expect gains beyond the generic good-for-all-cases codebook. That is, position-based learning opens up an opportunity to also adapt the beam codebook to the environment. In this paper, we propose an online beam pair selection and refinement algorithm to address the limitations of a completely offline approach. 

Our contributions are summarized as follows. 
\begin{itemize}
	\item We propose an online algorithm to learn to select beam pairs with risk-awareness to reduce the probability of severe beam misalignment during the learning. This is done by designing the algorithm to select high-risk beam pairs less often. The proposed solution balances the learning burden on early-stage users and the learning speed. 
	\item We provide regret analyses of the proposed algorithms, which provide insights 
	into the cost of the learning due to the introduction of risk-awareness. 
	\item We formulate the beam pair refinement problem as a continuum-armed bandit (CAB) problem. Our solution is based on the hierarchical optimistic optimization (HOO) \cite{Bubeck2011} with modifications to suit the beam alignment context. 
	\item We integrate the two algorithms into a two-layer online learning solution that learns to select and refine the beam pairs at the same time. The beam pair selection part learns coarse beam directions and the refinement part learns to refine them. This hierarchy is more efficient than learning the refined beam directions directly since now the refinement learning focuses only in promising directions selected by the beam selection part. 
\end{itemize}
Our numerical result in a vehicle-to-infrastructure (V2I) scenario shows that the integrated solution learns quickly. Using $16\times 16$ arrays at both the transmitter and receiver and a training budget of 30 beam pairs, it achieves an average gain of 1dB over the exhaustive beam search over $271^2$ beam pairs in the unrefined codebook within the first 100 time steps. The gain can reach up to about 1.5dB over time. 
Unlike prior work \cite{Seo2016,Scalabrin2018,Hashemi2018} that uses simplistic abstract models that match exactly with the underlying statistical assumptions of the problem formulation, we use realistic channels generated by ray-tracing to evaluate our algorithms. 


Beam alignment has been investigated intensively in the literature. Several directions have been pursued such as approaches based on beam sweeping \cite{Wang:Beam-codebook-based-beamforming-protocol:09,Hosoya:MIDC-a-novel-beamforming-technique:14}, angle of arrival and departure (AoA/AoD) estimation \cite{Duan2015,Marzi2016}, black-box function optimization \cite{Kadur2016,Li2013}, and the use of side-information \cite{Nitsche2015,Devoti2016,Aviles2016,Inverse-fingerprint-preprint}. 
We refer to \cite{Inverse-fingerprint-preprint} for a summary of the differences of these approaches for beam alignment. The last category is most related to our work, especially those that use position information.  

\redtext{
There are different approaches to position-aided beam alignment. 
One line of research assumes LOS channels and determines the pointing direction directly from the transmitter and receiver position \cite{Va:beam-design:2016,Garcia2016,Abbas2016,Junil2016}. That approach does not require beam training when the position information is accurate. Position error can be translated into uncertainty in the AoA/AoD. A small amount of beam training covering the AoA/AoD uncertainty range can counter the position error \cite{Junil2016,Garcia2016}. While efficient, the LOS assumption is not always valid. For example, in our ray-tracing simulations, the channel is NLOS about 38\% of the time. 

In another line of research, a database of past beam measurements indexed by location is used to identify both NLOS and LOS beamforming directions \cite{Inverse-fingerprint-preprint,Aviles2016,Devoti2016}.
The work in \cite{Devoti2016} proposed to store the most recent successful antenna configuration (having received power larger than some threshold) at each location (defined as grid points). The objective here is to find a direction that can support a link while our objective is to find the beam pair that provides the strongest received power. 
Also, omni-directional users are assumed in \cite{Devoti2016}, which limits the achievable communication range. 
A hierarchical search was proposed in \cite{Aviles2016} assuming users are equipped with horn antennas with different beamwidths. A hierarchical search requires multiple feedbacks, which could be the bottleneck as they are sent via a slow link in the control plane. Unlike \cite{Devoti2016,Aviles2016}, our prior work in \cite{Inverse-fingerprint-preprint} assumed directional users and performs the beam training at the beam level without hierarchical search requiring only one feedback. Also, the method in \cite{Inverse-fingerprint-preprint} uses the probability of being optimal (i.e., alignment probability) as the selection metric. This method can be shown to be optimal in maximizing the probability of finding the strongest beam direction \cite{Inverse-fingerprint-preprint}. In this paper, we propose an online version of this optimal selection method from \cite{Inverse-fingerprint-preprint}. We also develop a new beam pair refinement method to adapt the beam codebook to the environment to further maximize the beamforming gain. 

Another line of research uses position information of surrounding scatterers to further increase the beam alignment efficiency \cite{Wang2018,Maschietti2017}. In \cite{Wang2018}, a decision-tree algorithm was proposed for a mmWave V2I beam alignment using positions of neighboring vehicles as the features. The training dataset consists of past beam measurement and feature pairs. 
In a more traditional (i.e., not a data-driven approach leveraging past beam measurements) and abstract setting, the positions of dominant scatterers are assumed known with some error in \cite{Maschietti2017}, and a distributed beam alignment framework was proposed. While more information will be helpful, additional sensors and/or procedures are required to obtain those positions and they may be sensitive to the level of knowledge. Incorporating more side informaiton is an interesting direction for future work. In this paper, we will limit to using only the transmitter/receiver position, i.e., this work belongs in the category described in the previous paragraph. 
}

The recent progress in ML has revived interest in applying ML techniques to communications \cite{Jiang2017,Chen2017,Ye2018}. 
Related work that applies ML to beam alignment includes \cite{Va2017,Wang2018,Alkhateeb2018}. 
Position-aided beam prediction was proposed in \cite{Va2017,Wang2018}. Decision tree learning was used in \cite{Wang2018}, and a learning-to-rank method was used in \cite{Va2017}. 
A coordinated beamforming solution using deep learning was proposed in \cite{Alkhateeb2018}. Here, the received training signals via omni reception at a set of coordinating BSs are used as the input to a deep learning model that predicts the beamforming vectors at those BSs to serve a single user. 
The work in \cite{Va2017,Wang2018,Alkhateeb2018} shows that machine learning is valuable for mmWave beam prediction. Unfortunately, the proposed methods are all supervised learning techniques, which assume an offline learning setting and require a separate training data collection phase.
In this paper, we propose online learning algorithms using the multi-armed bandit (MAB) framework, which is a special class of reinforcement learning (RL). 

Recent applications of RL/MAB for beam training include \cite{Seo2016,Scalabrin2018} which uses a partially observable Markov decision process (POMDP) framework, and \cite{Gulati2014,Hashemi2018} which uses an MAB framework. 
The work in \cite{Seo2016,Scalabrin2018} addresses tracking problems where the POMDP with known state transition models provides a means to predict the state of the channel enabling an informed choice of the probing beams for good performance. The state transition models, however, are not easily obtained in a practical setting.  
In \cite{Hashemi2018}, the beam alignment problem is solved using an MAB framework with the assumption that the success probability (the received power is larger than some threshold) is a unimodal function of the pointing direction. The efficiency of that solution depends on this unimodal property, which cannot be guaranteed in our setting with random blockage. The work in \cite{Gulati2014} implemented the UCB1 algorithm and its variants from \cite{Auer2002} for antenna state selection for a 2.4GHz IEEE 802.11 system. Note that both \cite{Gulati2014} and \cite{Hashemi2018} proposed single-play MAB solution, while this paper proposes a multiple-play MAB. The two types of MAB will be described next. 

MAB is a useful tool for solving \redtext{ online learning (also called sequential decision-making)} problems \cite{Bubeck2012,Burtini2015}. The most common form of MAB is the single-play MAB with a finite number of arms, where only one arm is selected in each time step. The proposed method trains multiple beam pairs (up to the training budget) in each beam alignment attempt. 
Thus, our beam pair selection problem can be cast as a multiple-play MAB problem (also known as combinatorial bandit) \cite{Chen2013}, where multiple arms may be tried in each round or time step.
In an MAB setting, in each round, the player must decide between using the knowledge obtained so far to select the best arm or explore lesser-known arms, which is called the explore-exploit dilemma. The optimism in the face of uncertainty is a core design idea for balancing the explore-exploit tradeoff, which results in a widely successful family of algorithms known as the upper confidence bound (UCB). Our solution employs the UCB in a multiple-play setting. Most related to our solution is the cascading bandit \cite{Kveton2015}, which performs the same selection procedure as our Algorithm \ref{alg:greedy_UCB_algorithm} but with a different model to collect the reward measurements. Another important difference is that the reduction to the greedy selection is based on the independent arms assumption in \cite{Kveton2015}, while in our case it is based on the modularity property (i.e., additivity) of the reward signal as a function of the subset of selected arms. Also, we extend beyond the greedy UCB selection by introducing risk-awareness designed to avoid severe beam misalignment during the learning. 

We cast our beam pair refinement as a stochastic CAB problem, which has infinitely many arms. CAB assumes the reward function has some smoothness property (e.g., Lipschitz continuous). There are different approaches to solve CAB such as Bayesian optimization (BO) \cite{Srinivas2010}, the zooming algorithm \cite{Kleinberg2008}, and optimistic optimization (OO) \cite{Munos2014}. BO does not discretize the arm space but has high complexity. It is more suitable when sampling is expensive or the learning horizon is short. The zooming algorithm and OO rely on smart discretization of the arm space. The zooming algorithm uses an adaptive approach that applies finer discretization in promising regions. This is done using an activation rule that is assumed given to the algorithm, but this rule is a non-trivial problem itself. OO approaches discretize the arm space using a tree and exploit the hierarchy for an efficient search for the best arm. OO approaches designed for stochastic settings include Stochastic Simultaneous OO (StoSOO) and HOO \cite{Munos2014}. StoSOO is an explore-first algorithm where the task is to find the best arm given an exploration budget. This does not fit our setting where there is no separate explore and exploit phases. HOO is designed for maximizing the cumulative reward and suits our setting well. Applying HOO in its original form does not work well. We propose three modifications to suit our beam refinement problem. 



The rest of the paper is structured as follows. Section~\ref{sec:system_model} describes our system model and how we generate the data. Section~\ref{sec:beam_alignment_using_beam subset} presents our beam alignment framework and reviews offline beam pair selection methods from \cite{Inverse-fingerprint-preprint}, which are the basis for our online learning solution. 
Section~\ref{sec:two_layers_online_learning} describes the proposed two-layer online learning algorithm with the beam pair selection in the first layer and beam pair refinement in the second. 
Section \ref{sec:online_beam_pair_selection} and Section~\ref{sec:online_beam_refinement} provide the details of the two layers along with some analysis. 
Numerical evaluations are given in Section~\ref{sec:numerical_results} followed by the conclusions in Section~\ref{sec:conclusions}.  

\section{System and data model} \label{sec:system_model}
In this section, we describe our model of the communication system and how we generate the data for evaluating the learning algorithms. 

\subsection{System model} \label{sec:sys_model}
Our system model is comprised of the channel model, the received signal model, and the codebook. 
We assume a wideband geometric channel model that is widely used in mmWave simulations \cite{Heath2016}. We denote $N_\rmt$ and $N_\rmr$ the numbers of transmit and receive antennas, $L_\rmp$ the number of rays, $T$ the symbol period, $g(\cdot)$ the combined response of matched and lowpass filtering, $\ba_\rmt$ and $\ba_\rmr$ the normalized transmit and receive array steering vectors, $\alpha_\ell$ the complex channel gain, $\tau_\ell$ the delay, $\phi_\ell^\rmA$ and $\phi_\ell^\rmD$ the azimuth AoA and AoD, $\theta_\ell^\rmA$ and $\theta_\ell^\rmD$ the elevation AoA and AoD of the $\ell$-th path, and $(\cdot)^*$ the conjugate transpose operator. The channel at the delay tap $m$ is given by
\begin{align}
\label{eq:channel_model}
\bH[m] = \sqrt{ N_\rmr N_\rmt} \sum_{\ell=0}^{L_\mathrm{p}-1}\!\!  \alpha_\ell g(mT-\tau_\ell) \ba_\rmr(\theta_\ell^{\rmA},\phi_\ell^{\rmA})
\ba_\rmt^*(\theta_\ell^{\rmD},\phi_\ell^{\rmD}). 
\end{align}
We parametrize this channel using ray-tracing assuming single antennas at both the transmitter and receiver. This means $\alpha_\ell$ is the channel gain of a single antenna, and the factor $\sqrt{ N_\rmr N_\rmt}$ is needed to express the array gain (since $\ba_\rmt$ and $\ba_\rmr$ are normalized). 

We assume an analog beamforming architecture with one RF chain. Denote $L$ the channel length, $P_\rmt$ the transmit power, $s[k]$ the known training signal, $\bH_\ell = \sqrt{N_\rmr N_\rmt} \alpha_\ell \ba_\rmr(\theta_\ell^{\rmA},\phi_\ell^{\rmA}) \ba_\rmt^*(\theta_\ell^{\rmD},\phi_\ell^{\rmD})$ the channel matrix corresponding to the $\ell$-th path, $v_i[k]$ the complex Gaussian noise $\mathcal{CN}(0,\sigma_v^2)$, and $r(i)$ and $t(i)$ the mappings from the beam pair index $i$ to the combiner $\bw$ and beamformer~$\beff$ vector indices, the received signal 
is given by 
\begin{align}
& y_i[k] = \nonumber \\ 
&\sqrt{P_\rmt} \sum_{m=0}^{L-1}\!\! s[k-m] \underbrace{\sum_{\ell=0}^{L_\mathrm{p}-1} \!\! g(mT+\tau_0-\tau_\ell)\bw_{r(i)}^* \bH_\ell \beff_{t(i)}}_{h_i[m]} + v_i[k]. 
\label{eq:Rx_signal_model}
\end{align}
The channel strength is defined as the squared norm of $\bh_i =[h_i[0],\dots,h_i[L-1]]^\rmT$, i.e.,
\begin{align}
\gamma_i = \|\bh_i\|^2. 
\end{align}
The effective channel $\bh_i$ can be estimated from noisy signals, for example using a least-squared estimator \cite{wirelessLabTextbook}. The impact of noise on the beam alignment has been studied in \cite{Inverse-fingerprint-preprint,Va2018}, where it is shown that the impact is small for a wideband system. 
For a clear exposition of the learning algorithms, we assume noise-free $\gamma_i$ in this paper. 

The beam codebook used in this paper is generated using progressive phase-shift \cite{Balanis:Antenna-theory-analysis-and-design:05}. We note that this choice is not important, and other codebooks such as the DFT codebook could be used. Uniform planar arrays (UPA) are assumed at both the transmitter and the receiver. With a UPA, each beam is defined by its azimuth $\phi$ and elevation $\theta$ main beam direction. Let $G_\mathrm{ant}(\cdot)$ be the antenna element radiation pattern, $\Omega_\rmy = k d_\rmy\sin(\theta)\sin(\phi)$, $\Omega_\rmx = k d_\rmx\sin(\theta)\cos(\phi)$, $k=2\pi/\lambda$ be the wave number, $\otimes$ denote the Kronecker product, $N_\rmx$ and $N_\rmy$ be the numbers of elements along the x- and y-axis, and $d_\rmx$ and $d_\rmy$ be the element spacing in the x- and y-direction, a beam pointing in $(\theta,\phi)$ direction is given by \cite{Balanis:Antenna-theory-analysis-and-design:05}
\begin{align}
\ba(\theta,\phi) = \frac{G_\mathrm{ant}(\theta,\phi)}{\sqrt{N_\rmx N_\rmy}}
\begin{bmatrix}
1 \\
e^{\jj \Omega_\rmy}  \\
\vdots \\
e^{\jj (N_\rmy-1)\Omega_\rmy} 
\end{bmatrix} \otimes \begin{bmatrix}
1 \\
e^{\jj \Omega_\rmx} \\
\vdots \\
e^{\jj (N_\rmx-1)\Omega_\rmx}
\end{bmatrix}.
\end{align}
We assume $d_\rmx=d_\rmy=\lambda/2$ in this work. We assume no backplane radiation and set  
\begin{align}
\label{eq:ante_pattern}
G_\mathrm{ant}(\theta,\phi)=
\begin{cases}
1 & \text{if } \theta < 90^\circ \\
0 & \text{otherwise}
\end{cases}. 
\end{align}
We use \eqref{eq:ante_pattern} for simplicity, but we can replace it with a more sophisticated pattern like that of a patch antenna. 

\begin{figure}
\centering
\includegraphics[width=0.9\columnwidth]{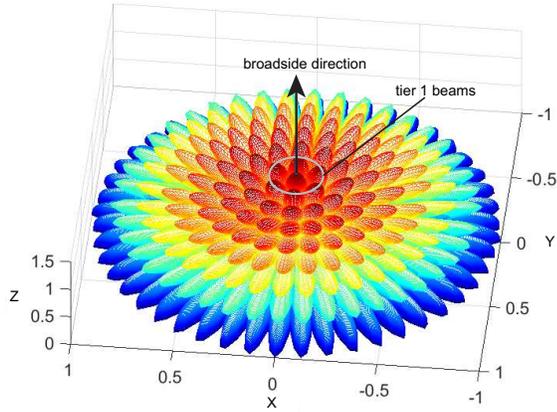}
\caption{Beam patterns in our codebook for an $8\times8$ array. The array is assumed to face upward in the $+z$ direction. The codebook covers the directions in the $+z$ half-space (i.e., assuming no radiation in the backplane).}
\label{fig:8by8_array_3Dpatterns}
\end{figure}

The beams are generated such that they are separated by their 3dB beamwidth in the azimuth and elevation. Fig. \ref{fig:8by8_array_3Dpatterns} shows the beams for an $8\times8$ array. The procedure starts from the broadside direction. First, fixing the azimuth angle at $0^\circ$, the elevation beam direction that crosses the broadside beam at the 3dB point is determined numerically. We then do the same procedure in the azimuth while fixing the elevation angle until all $360^\circ$ are covered (call these the tier 1 beams). Next, we compute the elevation beam direction that crosses the tier~1 beam at the 3dB beamwidth and determine all the azimuth directions until all $360^\circ$ are covered. This is repeated until the main beam direction in elevation exceeds $90^\circ$, i.e., reaching the backplane direction. 

\subsection{Data model} \label{sec:data_model}

\begin{figure}
\centering
\includegraphics[width=0.7\columnwidth]{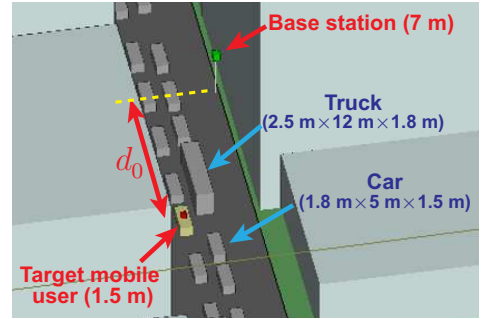}
\caption{A snapshot of the ray-tracing simulation in an urban street. The street has two lanes, and two types of vehicles (cars and trucks) are simulated. The BS's antenna is at 7m and the MU's antenna is at 1.5m from the ground.}
\label{fig:WI_sim_scenario}
\end{figure}

To generate the data, we parametrize the channel model \eqref{eq:channel_model} using a commercial ray-tracing simulator, called Wireless InSite \cite{Remcom_WI}. Ray-tracing ensures spatial consistency in the channels, 
which is essential to our location-based learning problems.
The ray-tracing simulation is shown in Fig. \ref{fig:WI_sim_scenario}. We assume a V2I setting in an urban street canyon.  The street has two lanes, and there are two types of vehicles represented by metal boxes: cars ($1.8\rmm\times5\rmm\times1.5\rmm$) and trucks ($2.5\rmm\times12\rmm\times3.8\rmm$). 
The mobile user (MU) is a car on the far side lane from the BS. 
Because it is larger, a truck can cause blockage to the MU. We assume roof-mounted MU antenna at 1.5m and BS antenna at 7m. We generate the channel in a per snapshot basis. In each snapshot, the MU is placed uniformly at random within a location bin $[d_0-\sigma_\rmd,d_0+\sigma_\rmd]$ in the far side lane, and all other vehicles are placed randomly with their gaps following an Erlang distribution. We refer to \cite{Inverse-fingerprint-preprint} for a detailed description including the material properties used for the ray-tracing. 
We set the carrier frequency to $60$GHz, $d_0=30$m, and $\sigma_\rmd=2.5$m, which corresponds to a location bin of $5$m length. Note that our method only requires that the position be accurate enough to identify the location bin. The system bandwidth is set to $1.76$GHz, which is used in IEEE 802.11ad. The system bandwidth is used to compute the symbol period used in the channel model \eqref{eq:channel_model}. Using these parameters, we generate 10,000 channel samples. We assume $16\times 16$ UPA at both the MU and the BS when computing the channel strengths. The codebook described in Section~\ref{sec:sys_model} has 271 beams.


\section{Position-aided beam alignment} \label{sec:beam_alignment_using_beam subset}

\begin{figure}
	\centering
	\includegraphics[width=0.8\columnwidth]{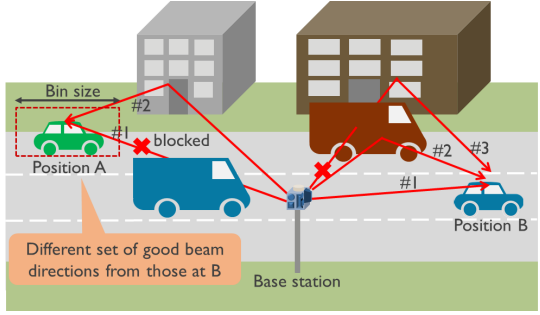}
	\caption{\redtext{Illustration of the intuition of the proposed position-aided beam alignment. Consider a vehicle at position A. The geometry of the environment only allows two possible pointing directions: the LOS and the building-reflection path. If the system can learn from past beam measurement results at position A to identify the two beam directions, then beam training can be reduced to just train these two directions. In an actual setting there will be position error. In our proposed solution, we use location bin that allows position inaccuracy to be in the range of the bin size. }}
	\label{fig:fingerprinting_intuition}
\end{figure}

Our beam alignment method relies on the premise that context information (e.g., position) can be used to reduce the training overhead. Using context, the method only needs to train the most promising beam directions. \redtext{Fig. \ref{fig:fingerprinting_intuition} illustrates the intuition behind the proposed method. Consider the vehicle at position A. In this case, the geometry of the environment allows only two pointing directions: the LOS path and the building reflection path. If the system can learn from past beam alignment experience to identify these two directions, the beam alignment overhead can be reduced to training only these two directions. Now, in practice there will be position error. The larger the error, the more uncertain the pointing directions, and the required beam training overhead will increase. In our simulations, we allow position inaccuracy to be within the 5m location bin, and a training budget of only 30 beam pairs is enough to achieve negligible performance loss compared to the exhaustive search when using $16\times 16$ arrays. (See \cite[Section VI.C]{Inverse-fingerprint-preprint} for a detailed evaluation of the effect of the location bin size.) Note that the edge effect at bin boundaries can be mitigated by defining overlapping bins. The small number of possible pointing direction given a location has also been observed in measurements. For example, 
\cite[p. 19]{Chizhik2017} reported that in a 28GHz indoor measurement setting with moving pedestrians the received powers were concentrated in no more than three dominant directions.
} 


In this section, we first describe the position-aided beam alignment framework. Then, we explain the beam alignment accuracy metric, based on which the core of the 
framework, the beam pair selection method, is developed. 
Lastly, we review the offline beam pair selection method from \cite{Inverse-fingerprint-preprint} which will be used in our online solution. 

\subsection{Overview of the proposed position-aided beam alignment}

\begin{figure}
	\centering
	\includegraphics[width=\columnwidth]{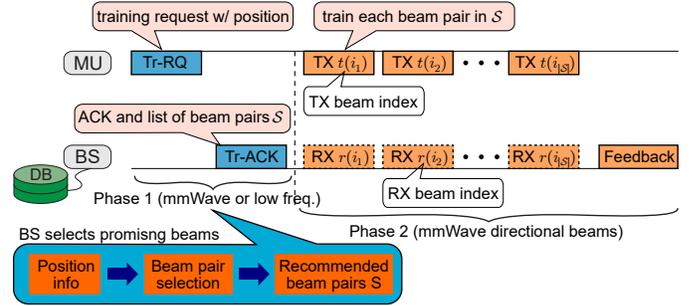}
	\caption{An illustration of position-aided beam alignment in the uplink. It consists of two phases. Phase 1 is for the training request where the MU position is sent to the BS. The BS uses the position and its learned database to determine a list of promising beam pairs $\mcS$. In Phase 2, the beam pairs in the list are trained, and a feedback indicating the best beam index is sent at the end. \redtext{The database used for the beam pair selection is stored and maintained at the BS without any burden on the MU.}}
	\label{fig:beam_alignment_using_beam_subset}
\end{figure}

\redtext{ 
The idea of our proposed approach is to learn from past beam measurements experienced in a given discretized location. The measurements themselves and/or learning parameters are stored and maintained at the BS. In the offline learning case as in our prior work \cite{Inverse-fingerprint-preprint}, a sufficient number of beam measurements are assumed collected before they are used to recommend beam pairs. In the online learning case considered in this paper, the learning parameters of a location bin are updated every time a beam alignment attempt is made and a new set of beam measurements becomes available at that location bin. 

Assuming the database is available for making recommendation, we describe how the beam alignment procedure works.}
Fig. \ref{fig:beam_alignment_using_beam_subset} illustrates the position-aided beam alignment, which consists of two phases. We start with the uplink. In Phase 1, the MU sends a training request along with its context information to the BS. In this work, we use position as the context. The BS uses the position and the database it maintains to determine a subset of promising beam directions, denoted by $\mcS$. The size of the set $\mcS$ is a system parameter chosen to balance the training overhead and the alignment accuracy. The BS then responds with an acknowledgment and the beam pair subset $\mcS$ to the MU. Since the beams are not aligned in this phase, a lower frequency control channel or mmWave with a large spreading factor can be used. 
In Phase 2, the beam pairs in $\mcS$ are trained and the best beam index is fed back at the end. MmWave with directional beams is used during this phase.  

In the downlink, Phase 1 changes slightly. The process starts with the BS sending a training request to the MU, which then responds with an acknowledgment including its position. Next, the BS sends the list of promising beam pairs $\mcS$. The beam training in Phase 2 is kept the same. 
This is possible because of the reciprocity in the AoAs/AoDs, where the AoAs become the AoDs and vice versa when reversing the transmitter and receiver role. This AoA/AoD reciprocity only depends on the reciprocity property of electromagnetic waves, which holds when they propagate in passive medium like wireless channels (excluding the device's circuits)  \cite{Balanis:Antenna-theory-analysis-and-design:05}. 

\redtext{Finally, we emphasize that this beam alignment framework is flexible and different beam pair selection methods can be used for the ``Beam pair selection" block. In particular, we proposed an offline learning approach in \cite{Inverse-fingerprint-preprint} (briefly described in Section \ref{sec:offline_inv_fingerprint}) that assumes the past beam measurements are already collected. In this paper, we propose online algorithms to select the beam pairs, which will be detailed in Section \ref{sec:two_layers_online_learning}.} 


\subsection{Quantifying beam alignment accuracy}
There are many ways to quantify the accuracy of beam alignment. Here, we use the power loss and the power loss probability. The power loss can capture the severity of the misalignment and differentiate whether the current alignment is 3dB or 10dB away from the optimal alignment. This is important because while both cases are misaligned, the former likely still provides a good link while the latter likely cannot.   

We start with the definition of the power loss. It is the ratio of the channel strength between the selected beam pair indexed by $s$ and the optimal channel strength given by
\begin{align}
\label{eq:powLoss_def}
\xi = \frac{\max_{i\in\mcB} \gamma_i}{\gamma_s},
\end{align}
where $\mcB$ denotes the set of all possible beam pairs in the codebook. The beam pair $s$ is selected from the selection set $\mcS$, and with accurate beam training $s=\arg\max_{i\in\mcS}\gamma_i$. If the codebook is used without any modification, then $\mcS\subset \mcB$ and $\xi\ge 1$ always holds. The proposed online learning method also includes a component to refine the beam pairs to adapt the codebook to the environment, in which case $\xi<1$ is possible.  

We quantify the beam alignment accuracy by the power loss probability defined~by
\begin{align}
\label{eq:powLossProb_def}
P_\rmpl(c,\mcS) = \bbP\left[\xi > c \right],
\end{align}
for some constant $c\ge 1$. We call the case when $c=1$ the misalignment probability. A related concept to the misalignment probability is the probability of being optimal, given~by
\begin{align}
P_\rmopt(\mcS) = \bbP[i^\star \in \mcS],
\end{align}
where $i^\star=\arg\max_{i\in\mcB}\gamma_i$ denotes the index of the optimal beam pair. We note that 
\begin{align}
\label{eq:rel_powLossProb_probOpt1}
P_\rmopt(\mcS) & = \bbP[\xi=1] \\
\label{eq:rel_powLossProb_probOpt2} 
& = 1-\bbP[\xi>1] \\
\label{eq:rel_powLossProb_probOpt3}
& = 1- P_\rmpl(1,\mcS),
\end{align}
where \eqref{eq:rel_powLossProb_probOpt2} follows because $\xi\ge 1$ (without refinement). 
$P_\rmopt(\mcS)$ is used in the optimal beam selection method described in the next subsection.

\subsection{Offline inverse fingerprinting beam pair selection} \label{sec:offline_inv_fingerprint}
In this subsection, we review two offline beam pair selection methods from \cite{Inverse-fingerprint-preprint} which form the basis for our online solution. We begin by describing the offline database. It is assumed that this database has already been collected before it is used for beam pair selection. The database has $N$ observations, where each observation consists of the channel strengths of all beam pairs combinations in $\mcB$ measured in a location bin. It is assumed that each observation is measured within a beam coherence time so that the spatial channel has negligible change \cite{Va:Impact-of-beamwidth:2016}. An example of the database is shown in Table \ref{tab:raw_fingerprint_example}. We note that some tradeoff between the performance and the cost for collecting and storing the database is possible (see \cite{Inverse-fingerprint-preprint} for details). 

\begin{table}
\centering
\caption{An example of the database at a location bin. Each row consists of the measured channel strengths of all beam pair combinations for a given channel realization. 
Here, we denote $I$ the beam pair index and $\gamma$ the channel strengths in $\mathrm{dB}$. 
}
\label{tab:raw_fingerprint_example} 
\begin{tabular}{|c||>{\columncolor[gray]{0.7}}c|>{\columncolor[gray]{0.9}}c|>{\columncolor[gray]{0.7}}c|>{\columncolor[gray]{0.9}}c|c|>{\columncolor[gray]{0.7}}c|>{\columncolor[gray]{0.9}}c|} 
\hline 
\multirow{2}{*}{Obsv. No.} & \multicolumn{2}{c|}{Best} & \multicolumn{2}{|c|}{2nd best} & $\dots$ & \multicolumn{2}{c|}{$|\mcB|$-th best} \\ \hhline{~-------} 
 & $I$ & $\gamma$ & $I$ & $\gamma$ & $\dots$ & $I$ & $\gamma$ \\  \hline 
 1 & 5  & -64.5 & 159 & -69.2 & $\dots$ & 346 & -95.8 \\ \hline 
2 &  159& -70.4  & 263 & -72.6 & $\dots$ & 354 & -97.1 \\ \hline
    $\dots$ & $\dots$ & $\dots$ & $\dots$ & $\dots$ & $\dots$ & $\dots$ & $\dots$ \\ \hline 
$N$ &  5& -66.4  & 258 & -68.1 & $\dots$ & 2 & -82.6 \\ \hline
\end{tabular}
\end{table}

We now describe the two beam pair selection methods from \cite{Inverse-fingerprint-preprint}, namely AvgPow and MinMisProb. AvgPow is a heuristic that selects the beam pairs by their average channel strengths. Denote $\bar{\gamma}_i$ the sample average of the channel strength of the $i$-th beam pair and $\underset{i\in\mcB; M}{\arg\max}\{\cdot\}$ the operator that returns the top-$M$ indices, the selection set of size $|\mcS_\mathrm{AP}|=M$ can be written~as
\begin{align}
\label{eq:AvgPow_selection}
\mcS_\mathrm{AP} = \underset{i\in\mcB; M}{\arg\max} \left\{ \bar{\gamma}_i \right\}. 
\end{align}
MinMisProb is an optimal selection method that minimizes the misalignment probability. 
It can be shown that this optimal selection reduces to a greedy selection using the probability of being optimal \cite{Inverse-fingerprint-preprint}. 
Let $|\mcS_\mathrm{MMP}|=M$ and denote $\hat{P}_\rmopt(i)$ the probability of being optimal of the beam pair $i$ estimated from the database, then
\begin{align}
\label{eq:MinMisProb_selection}
\mcS_\mathrm{MMP} = \underset{i\in\mcB; M}{\arg\max} \left\{ \hat{P}_\rmopt(i) \right\}.
\end{align}
In this paper, we develop a risk-aware online learning version of MinMisProb, where beam pairs selected by the UCB indices are rejected with a probability reflecting their risk of low received powers. 
When rejected, the replacement pair is selected using both MinMisProb and AvgPow. 


\section{Proposed two-layer online learning} \label{sec:two_layers_online_learning}
Our aim in this paper is to develop an online learning algorithm for fast and efficient beam alignment. 
We propose a two-layer online solution to achieve this goal. 
The idea here is to learn coarse beam directions (quantized by the 3dB beamwidths) that are promising in the first layer and conduct a refinement of those promising directions in the second layer. This kind of hierarchy is efficient because the refinement is only done in promising directions. 

\begin{figure}
\centering
\includegraphics[width=0.9\columnwidth]{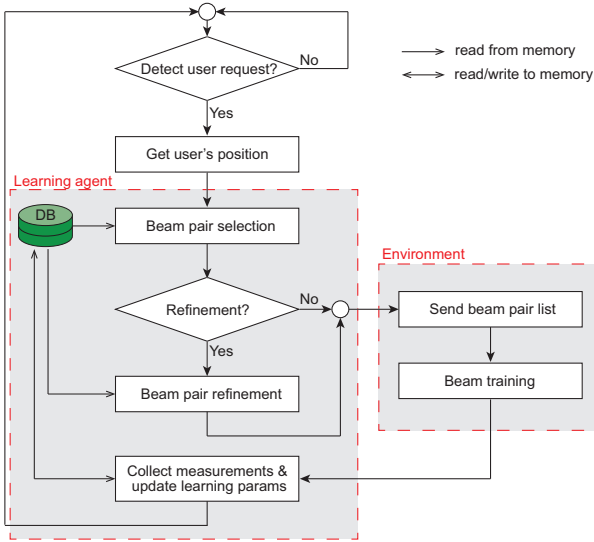}
\caption{A flowchart of the two-layer online learning. The algorithm starts with a training request detection loop. When it detects a request, the algorithm decodes the user's position and input to the beam selection procedure, which then reads the learning parameters corresponding to the position and determines a subset of promising beam pairs. If beam refinement is enabled, the refinement parameters of those selected pairs are selected. The beam subset is then sent to the user and the subset of beam pairs are trained. The beam measurements are used to update the learning parameters and the algorithm returns to the training request detection loop.}
\label{fig:overview_two_layer_learning}
\end{figure}

An overview of the proposed online learning solution is illustrated in Fig.  \ref{fig:overview_two_layer_learning}. \redtext{Note that this is a solution for the ``Beam pair selection'' block in Fig. \ref{fig:beam_alignment_using_beam_subset}.} The learning happens at the BS as explained earlier. The algorithm runs in an infinite loop, where in each iteration, it recommends a list of beam pairs and updates the learning parameters recorded in the database upon receiving the beam measurements of those pairs. As mentioned in Section \ref{sec:offline_inv_fingerprint}, by having the MU transmit, there is no extra feedback overhead to collect the beam measurements. We highlight the groups of blocks that correspond to the learning agent and the environment in Fig. \ref{fig:overview_two_layer_learning}. This shows a typical RL setting where the agent optimizes its action through direct interaction with the environment \cite{Sutton1998}. In our problem, the action is the subset of beam pairs selected for the training, and the environmental response is the beam measurement results. The algorithm starts by running a detection loop for a request for beam training from the user. If a request is detected, the position (other context can also be used, but we focus on position) is extracted from the training request packet and input to the beam pair selection procedure. 
Then, the procedure produces a list of beam pairs using the learning parameters corresponding to the location bin stored in the database. 
If the beam pair refinement is enabled, the refinement parameters of the selected beam pairs are picked by the beam pair refinement procedure. The resulting subset of beam pairs is then sent to the user with an acknowledgment to allow the beam training. After beam training, the measurements of the selected beam pairs are used to update the learning parameters in the database. Then, the algorithm goes back to the detection loop to wait for the next training request. 


\section{Online beam pair selection} \label{sec:online_beam_pair_selection}

In this section, we describe the first layer of the two-layer solution. We start with the problem statement. Then, we develop online beam pair selection algorithms, first without and then with risk-awareness. We conclude the section with regret analyses of the proposed algorithms and some discussion.  

\subsection{Problem statement} \label{sec:prob_stat_beam_selection}
Our goal here is to develop an online version of the optimal beam pair selection method, MinMisProb. Specifically, the algorithm needs to solve the following optimization problem in an online setting: 
\begin{align}
\begin{array}{rl}
\underset{\mcS}{\mathrm{minimize}} & P_\rmpl(1,\mcS) \\
\mathrm{subject\,\, to} & \mcS \subset \mcB, \,\, |\mcS| \le \Btr,
\end{array}
\label{eq:subset_sel_prob}
\end{align}
where $\Btr$ is the desired subset size (the training budget). 
In an online learning setting, $P_\rmpl(1,\mcS)$ is not known, and it has to be estimated on the fly. 
To gain accurate knowledge of the beam pairs, each of them must be sampled multiple times, which means the learning can be very slow when $\mcB$ is a large set (i.e., when using large arrays with narrow beams). To remedy this problem, we propose to apply a heuristic to screen the beam pairs using a small offline database (of size $N$) to obtain a smaller set $\hat{\mcB}$ to apply the learning algorithm on. 
$\hat{\mcB}$ is obtained as the set of the unique beam pairs among the $NC$ entries of the first $C$ columns of Table \ref{tab:raw_fingerprint_example}. In our simulations, the offline database size $N=5$ and $C=200$ seem to be good enough for this purpose. 

\subsection{Greedy UCB algorithm}
We first propose a solution to \eqref{eq:subset_sel_prob} without risk-awareness. 
A subset $\mcS$ can be treated as a super-arm, and a single-play MAB algorithm can be used. Such an approach is not efficient because it treats each super-arm as independent and the number of super-arms is large due to the combinatorial nature of the number of all possible subsets. 

A more efficient approach to solve \eqref{eq:subset_sel_prob} is to leverage the structure of $P_\rmpl(1,\mcS)$ to take advantage of the dependence between the subsets. Specifically, we make use of the modularity property of the probability of being optimal \cite{Inverse-fingerprint-preprint}. We note that by the relationship in \eqref{eq:rel_powLossProb_probOpt3}, the problem \eqref{eq:subset_sel_prob} is equivalent to a maximization of $P_\rmopt(\mcS)$ with the same constraints, i.e.,
\begin{align}
\begin{array}{rl}
\underset{\mcS}{\mathrm{maximize}} & P_\rmopt(\mcS) \\
\mathrm{subject\,\, to} & \mcS \subset \hat{\mcB}, \,\, |\mcS| \le \Btr,
\end{array}
\label{eq:subset_sel_prob_equiv}
\end{align} 
where we also replace $\mcB$ by $\hat{\mcB}$ as explained earlier. 
Since $P_\rmopt(\mcS)$ is modular \cite{Inverse-fingerprint-preprint}, it can be decomposed as
\begin{align}
P_\rmopt(\mcS) = \sum_{i\in\mcS} P_\rmopt(i).
\label{eq:modularity_prob_opt}
\end{align} 
This property is due to the exclusive nature of the events that the $i$-th beam pair is optimal (i.e., having the highest channel strength). Recall that the probability of a union of exclusive events is the sum of the probability of each individual event \cite{Grimmett:probability-and-random-processes:01}. 
The main implication of \eqref{eq:modularity_prob_opt} is that the reward of $\mcS$ can be computed from the individual rewards of each of the beam pairs in $\mcS$. This means the optimal beam pair subset can be obtained by a greedy approach, where one beam pair is selected at a time. 
Observing this property, we propose to use a greedy UCB algorithm as shown in Algorithm~\ref{alg:greedy_UCB_algorithm}, that selects the beam pairs greedily using their UCB indices. The UCB index of an arm is a high confidence bound of the expected reward, which consists of the expected reward seen so far and the uncertainty (the confidence margin) \cite{Auer2002}. 

{\renewcommand{\baselinestretch}{1.0}
\begin{algorithm}
 \caption{Greedy UCB}
 \label{alg:greedy_UCB_algorithm}
 \begin{algorithmic}[1]
  \STATE // initialize arms' parameters using a small offline database 
  \STATE $X_\mathrm{tot}[i] \leftarrow 0$, for $\forall i\in \hat{\mcB}$ 
  \STATE $X_\mathrm{tot}[\arg\max_{i\in\hat{\mcB}}\bar{\gamma}^{\mathrm{init}}_i] \leftarrow 1$ 
  \STATE $T_i \leftarrow 1$, for $\forall i\in \hat{\mcB}$
  \FOR {$n=1,2,\dots$} 
  \STATE // Compute UCB values 
  \STATE $\UCB_i \leftarrow \frac{X_\mathrm{tot}[i]}{T_i}+ \sqrt{\frac{2\log(n)}{T_i}}$, for $\forall i\in \hat{\mcB}$ 
  \STATE // Greedy selection using UCB values 
  \STATE $\mcS \leftarrow \emptyset$ 
  \FOR { $k=1,2,\dots, \Btr$ } 
  \STATE $\mcS \leftarrow \mcS \cup \underset{i\in\hat{\mcB}\setminus\mcS}{\arg\max}\, \UCB_i $ 
  \ENDFOR 
  \STATE Train the selected $\Btr$ beam pairs to get $\gamma_{i,n}$ for $\forall i\in\mcS$\\
  \STATE // Update the learning parameters \\
  \STATE $T_i \leftarrow T_i + 1$, for $\forall i\in\mcS$ \\
  \STATE $X_\mathrm{tot} [\arg\max_{k\in\mcS}\gamma_{k,n}] \leftarrow X_\mathrm{tot} [\arg\max_{k\in\mcS}\gamma_{k,n}]+1$ \\
  \ENDFOR
 \end{algorithmic} 
\end{algorithm}
}

An important component of Algorithm~\ref{alg:greedy_UCB_algorithm} is the reward signal. 
Since the expected reward is the probability of being optimal, an ideal choice for the reward signal is 
\begin{align}
\label{eq:true_reward_def}
x_{i,t} = \begin{cases}
1 & \text{if } i \text{ was best in } \hat{\mcB} \\
0 & \text{otherwise}
\end{cases}, 
\end{align} 
which takes the value 1 if the pair $i$ is best and 0 otherwise. 
In an actual setting, it is not known if a pair is the best in $\hat{\mcB}$ since only the beam measurements for the beam pairs in the subset $\mcS\subset\hat{\mcB}$ are available. The best guess would be the strongest pair among the beam pairs trained. Considering this limitation, we propose to use an alternative and practical reward signal, 
\begin{align}
\label{eq:alternative_reward_def}
x_{i,t} = \begin{cases}
1 & \text{if } i \text{ was best in } \mcS \\
0 & \text{all other pairs in }\mcS 
\end{cases}, 
\end{align}
which takes the value 1 for the pair with the strongest beam measurement in $\mcS$ and 0 for all other pairs in $\mcS$.  
Denoting $X_{\mathrm{tot}}[i]=\sum_{t=1}^{n}x_{i,t}$, the expected reward of beam pair $i$ at time $n$ is estimated by $\hat{P}_\rmopt(i)=X_\mathrm{tot}[i]/T_i$, where $T_i$ is the number of times that the pair $i$ was selected up to time $n$. 

An intuitive understanding of this alternative reward definition can be drawn from an analogy to a sport tournament. In each round, the winners from each subgroup from the previous round play against each other to decide who will proceed to the next round, which eventually will reach the championship. We, thus, expect that over time only strong beam pairs will receive rewards of 1. 
We believe that under certain assumption on the underlying reward statistics of the beam pairs, it is possible to provide some guarantee that $\hat{P}_\rmopt(i)$ will converge to the true $P_\rmopt(i)$ as $T_i\to\infty$. This is outside the focus of this paper and is left for future work. 

As will be seen in Section \ref{sec:numRes_beam_sel},
Algorithm~\ref{alg:greedy_UCB_algorithm} does not perform well. 
The main reason for this is because it only tries to minimize the cumulative regret and is oblivious to the multiple-play setting in the beam alignment problem. 
Since multiple beam pairs are trained, the subset $\mcS$ can be divided into two parts. One part is for exploitation that uses the knowledge obtained so far to select the beam pairs and the other part is for exploration that aims at improving the accuracy of the learning parameters. 
By balancing these two parts,
it is possible to reduce the risk (large power loss events) at any given round. In other words, in the multiple-play setting, the risk of large losses can be traded off with the speed of learning (time to get accurate statistics of the arms). 
Another point for improvement in Algorithm \ref{alg:greedy_UCB_algorithm} is that it throws away the magnitude information because the reward signal is binary.
Recall that the binary reward signal is needed because we make use of the modularity of $P_\rmopt(\mcS)$ that allows the greedy selection using the UCB indices. 
To remedy these weaknesses, we propose a risk-aware version of Algorithm~\ref{alg:greedy_UCB_algorithm}. 

\subsection{Risk-aware greedy UCB algorithm}

We first start with the definition of risk. 
A possible choice for the risk is the power loss, which measures the misalignment severity. 
Since only the beam pairs in $\mcS$ are trained, the channel strength of the optimal beam pair is not necessarily known (especially, during the early stage of the learning) and the power loss cannot be computed directly. Another important point is that this risk needs to be estimated. Therefore, it is crucial to quantify the uncertainty of the risk estimate for it to be useful for the beam pair selection.
For these two reasons, we propose to use a binary risk signal defined in terms of the ratio of the channel strength of the beam pair and the best beam pair in $\mcS$, i.e., the risk signal of the beam pair $i$ at time $t$ is given~by
\begin{align}
\label{eq:risk_signal}
z_{i,t}=\begin{cases}
1 & \text{if }\frac{\max_{k\in\mcS}\gamma_{k,t}}{\gamma_{i,t}}>\Grisk \\
0 & \text{otherwise} 
\end{cases},
\end{align} 
where $\Grisk$ is a risk threshold. The choice of $\Grisk$ will be discussed in Section \ref{sec:numRes_beam_refinement}. 

A way to capture the uncertainty is to put a prior distribution on the risk based on the observations seen so far (i.e., a Bayesian approach). 
By the definition \eqref{eq:risk_signal}, $z_{i,t}$ is Bernoulli distributed with some unknown parameter $\zeta$. It is well-known that the Beta distribution is the conjugate prior to the Bernoulli distribution \cite{JordanLectureNoteConjugatePriors}. This means that the belief on the risk of the beam pair $i$ upon seeing the measurements up to time $n$ can be updated conveniently by updating the parameters of the Beta distribution. Specifically, denoting $Z_\mathrm{tot}[i]=\sum_{t=1}^{n}z_{i,t}$ the prior is updated as
\begin{align}
\label{eq:Beta_risk}
\tilde{Z}_{n} \sim \mathsf{Beta}\left( 1+Z_\mathrm{tot}[i], 1+T_i-Z_\mathrm{tot}[i] \right).
\end{align}
Here, we assume that at time 0 without any observation, $\tilde{Z}_0 \sim \mathsf{Beta}(1,1)$, which is the uniform distribution over $[0,1]$. This is a reasonable assumption since no information on the beam pair~$i$ is available at time 0.   

{\renewcommand{\baselinestretch}{1.0}
\begin{algorithm}
 \caption{Risk-Aware Greedy UCB}
 \label{alg:risk_aware_greedy_UCB_algorithm}
 \begin{algorithmic}[1]
  \STATE // initialize arms' parameters using a small offline database 
  \STATE $X_\mathrm{tot}[i]\leftarrow 0$, for $\forall i\in \hat{\mcB}$ 
  \STATE $X_\mathrm{tot} [\arg\max_{i\in\hat{\mcB}}\bar{\gamma}^{\mathrm{init}}_i] \leftarrow 1$
  \STATE $Z_\mathrm{tot}[i] \leftarrow 0$, for $\forall i\in \hat{\mcB}$
  \STATE $T_i \leftarrow 1$, for $\forall i\in \hat{\mcB}$
  \FOR {$n=1,2,\dots$} 
  \STATE // Compute UCB values 
  \STATE $\UCB_i \leftarrow \frac{X_\mathrm{tot}[i] }{T_i}+  \sqrt{\frac{2\log(n)}{T_i}}$, for $\forall i\in \hat{\mcB}$ 
  \STATE // Greedy selection using UCB values 
  \STATE $\mcS \leftarrow \emptyset$ 
  \FOR { $k=1,2,\dots, \Btr$ } 
  \STATE $\ell \leftarrow \underset{i\in\hat{\mcB}\setminus\mcS}{\arg\max}\,\, \UCB_i $
  \STATE $\tilde{Z}_n\sim  \mathsf{Beta}\left( 1+Z_\mathrm{tot}[\ell], 1+T_\ell - Z_\mathrm{tot}[\ell] \right)$ \label{Alg2:Beta_risk} 
  \STATE $Z_n \leftarrow \tilde{Z}_n \times \sqrt{\frac{2\log(n)}{T_\ell }}$ if $X_\mathrm{tot}[\ell]>0$, else $Z_n \leftarrow \tilde{Z}_n $
  \STATE $\mathrm{Rej}\sim \mathsf{Ber}(Z_n)$
  \IF { $\mathrm{Rej}=0$ } 
  \STATE $\mcS \leftarrow \mcS \cup \{\ell\}$
  \ELSE {}
  	\IF { $\exists i \in \hat{\mcB}\setminus \mcS$ with $X_\mathrm{tot}[i]>0$ } \label{Alg2:alternative_sel_start}
  		\STATE $\mcS \leftarrow \mcS \cup \underset{i\in\hat{\mcB}\setminus\mcS}{\arg\max}\, \hat{P}_\rmopt(i) $
  	\ELSE {}
  		\STATE $\mcS \leftarrow \mcS \cup \underset{i\in\hat{\mcB}\setminus\mcS}{\arg\max}\, \bar{\gamma}_i $ 
  	\ENDIF \label{Alg2:alternative_sel_end}
  \ENDIF  
  \ENDFOR 
  \STATE Train the selected $\Btr$ beam pairs to get $\gamma_{i,n}$ for $\forall i\in\mcS$\\
  \STATE // Update the learning parameters \\
  \STATE $T_i \leftarrow T_i + 1$, for $\forall i\in\mcS$ \\
  \STATE $X_\mathrm{tot}[\arg\max_{k\in\mcS}\gamma_{k,n}] \leftarrow X_\mathrm{tot}[\arg\max_{k\in\mcS}\gamma_{k,n}]+1$ \\
  \STATE $Z_{\mathrm{tot}}[i] \leftarrow Z_{\mathrm{tot}}[i]+1$ if $(\max_{k\in\mcS} \gamma_{k,n})/\gamma_{i,n}>\Grisk$, $\forall i\in\mcS$
  \ENDFOR
 \end{algorithmic} 
\end{algorithm}
}

We next explain how the risk estimate along with the prior are used in the rejection mechanism to reduce the probability of large power loss events during the learning. 
The new algorithm is shown in Algorithm~\ref{alg:risk_aware_greedy_UCB_algorithm}, which we call the risk-aware greedy UCB algorithm. 
The new addition to Algorithm \ref{alg:greedy_UCB_algorithm} is the risk-aware feature that rejects a beam pair selected by the greedy UCB with a probability reflecting its risk. 
The rejection probability is determined using the risk drawn from the prior distribution given in \eqref{eq:Beta_risk} and the confidence margin in two steps. 
First, a random variable $\tilde{Z}_n$ is drawn from this prior (line \ref{Alg2:Beta_risk}). 
Then, it is multiplied by the confidence margin for those beam pairs with $X_\mathrm{tot}[\ell]>0$. The obtained $Z_n$ is the rejection probability. The second step is needed because any beam pair is subject to blockage and their risks are not zero. This means that if $\tilde{Z}_n$ is used directly as the rejection probability, even good beam pairs will be rejected with non-zero probability even when $n\to \infty$. The second step ensures the algorithm accepts the UCB selection for ``good" beam pairs with increasing probability over time.

The proposed rejection mechanism is a random method that rejects the beam pair with a probability $Z_n$. First, the algorithm draws a Bernoulli random variable $\mathsf{Rej}$ with parameter $Z_n$. 
If $\mathsf{Rej}=0$, the algorithm accepts the beam pair, otherwise it rejects the pair. 
In that case, the replacement beam pair is selected using $\hat{P}_\rmopt(i)$ when there are pairs with $X_{\mathrm{tot}}[i]>0$, and using the average channel strength $\bar{\gamma}_i$ when all remaining pairs have $X_{\mathrm{tot}}[i]=0$. 
Unlike Algorithm \ref{alg:greedy_UCB_algorithm}, which does not use the amplitudes of the beam measurements $\gamma_{i,n}$, here they are used to update the risk parameters and also used for the replacement selection. This new algorithm makes a fuller use of the measurement information as compared to Algorithm~\ref{alg:greedy_UCB_algorithm}.

\subsection{Regret analysis}

In this subsection, we derive regret bounds of the two algorithms that will provide insights on the effect of the rejection mechanism we introduced in Algorithm~\ref{alg:risk_aware_greedy_UCB_algorithm}. We make a few simplifications to the problems to allow tractable analysis which we will describe in detail when presenting the results. 
Proofs are provided in Appendix \ref{Apd:proof_UCB_greedy} and \ref{Apd:proof_risk_aware_UCB_greedy}. 

Before stating the results, we first describe the metric used for the evaluation. For this type of online learning problem, a widely used metric is the cumulative regret. It is defined as {\it the cumulative performance loss as compared to the performance of an oracle that always plays the best arm} \cite{Bubeck2012}. 
Translating this to the beam alignment problem, {\it the regret incurred in a time step is non-zero when the algorithm does not select the best subset of beam pairs $\mcS^\star$}. 
Assuming $|\mcS^\star|=\Btr$, the optimal selection in \eqref{eq:MinMisProb_selection} tells us 
that $\mcS^\star$ contains the top-$\Btr$ beam pairs with the highest probability of being optimal $P_\rmopt(\cdot)$. Now, we call the beam pairs with the $\Btr$-highest $P_\rmopt(\cdot)$ as optimal and the rest of beam pairs as suboptimal. Then, the cumulative regret increases whenever one or more suboptimal beam pairs are selected in the selection set~$\mcS$. 

In the following, we present what is called a problem-dependent bound on cumulative regret (we drop `cumulative' from now on for convenience), which quantifies the regret in terms of the optimality gap. The optimality gap is defined as the difference in the probability of being optimal for an optimal pair $i^\star$ and a suboptimal pair $\ell$, i.e.,
\begin{align}
\label{eq:optimality_gap_def}
\Delta_{\ell,i^\star} = P_\rmopt(i^\star)-P_\rmopt(\ell).
\end{align}
By definition, $0< \Delta_{\ell,i^\star} < 1$ if $P_\rmopt(i^\star)>0$. Note that $\Delta_{\ell,i^\star}$ measures the difficulty in discriminating the suboptimal pair $\ell$ from the optimal pair $i^\star$ for the particular problem at hand; thus, the name problem-dependent when the regret bound is expressed using optimality gaps.

For Algorithm \ref{alg:greedy_UCB_algorithm}, we assume the reward signal during the learning is the ideal reward and not the alternative one, i.e., we assume the reward signal is given by \eqref{eq:true_reward_def} instead of \eqref{eq:alternative_reward_def}. 
We make this assumption because it is intractable to deal directly with the dynamics of the alternative reward signal in \eqref{eq:alternative_reward_def}. The main step in deriving the regret bound is the application of the Chernoff-Hoeffding inequality to bound the probability that the sample average of the reward is within the UCB value. To apply the Chernoff-Hoeffding inequality, it is required that the sample rewards are IID, which cannot be guaranteed when using the alternative reward definition because its distribution depends on the history of the selection done so far. This, however, is a reasonable assumption, since we expect that \eqref{eq:alternative_reward_def} will approach \eqref{eq:true_reward_def} for large $n$, which is the domain where the regret bound is meaningful. 

\begin{theorem}
\label{thm:regret_bound_greedy_UCB}
Assuming that the ideal reward signal \eqref{eq:true_reward_def} is accessible, the expected regret at time $n$ of the greedy UCB algorithm is upper bounded by
\begin{align}
\label{eq:regret_bound_greedy_UCB}
R_1[n] \le & 8\log(n) \sum_{\ell\in\mcB\setminus\mcS^\star} \sum_{i^\star\in\mcS^\star} \frac{1}{\Delta_{\ell,i^\star}} \nonumber \\
& +\left( 1+\frac{\pi^2}{3} \right) \sum_{\ell\in\mcB\setminus\mcS^\star} \sum_{i^\star\in\mcS^\star} \Delta_{\ell,i^\star}. 
\end{align}
\end{theorem}

Theorem \ref{thm:regret_bound_greedy_UCB} shows a regret bound for Algorithm \ref{alg:greedy_UCB_algorithm}. 
The first term, which increases with $n$, dominates the bound. It increases as $\Delta_{\ell,i^\star}$ decreases. This makes sense because a small $\Delta_{\ell,i^\star}$ means more samples are needed to differentiate the suboptimal pair $\ell$ from the optimal pair $i^\star$ with confidence. 
The regret bound is $\mcO(\log(n))$, which is known to be optimal up to the constant coefficient in front of $\log(n)$ \cite{Bubeck2012}. This confirms that the algorithm is a reasonable solution. 

For Algorithm \ref{alg:risk_aware_greedy_UCB_algorithm}, we make two additional assumptions besides the accessibility of the ideal reward signal. The first assumption is that the rejection probability of any beam pair $\ell$ is constant, denoted by $1-\zeta_{\ell}$. This is used because the rejection probability of the algorithm is dynamic (depending on the observation so far) and is not tractable. With a large enough $n$, we expect the risk estimate to stabilize, and thus this is not an unreasonable assumption. The second assumption is that when rejected the replacement selection has an optimality gap $\tilde{\Delta}_{\ell,i^\star}$. 

\begin{theorem}
\label{thm:regret_bound_risk_aware_greedy_UCB}
Assuming that the ideal reward signal is available, the rejection probability of beam pair $\ell$ is $1-\zeta_{\ell}$, and that when rejected the optimality gap of the replacement selection is $\tilde{\Delta}_{\ell,i^\star}$, then the expected regret at time $n$ of the risk-aware greedy UCB algorithm is bounded by
\begin{align}
& R_2[n]  \le \frac{8\log(n)}{\delta^2} \sum_{\ell\in\mcB\setminus\mcS^\star} \sum_{i^\star\in\mcS^\star} \frac{1}{\Delta_{\ell,i^\star}}
\nonumber \\
& + \frac{8\log(n)}{\delta^2} \sum_{\ell\in\mcB\setminus\mcS^\star} \sum_{i^\star\in\mcS^\star} \frac{(1-\zeta_\ell)\tilde{\Delta}_{\ell,i^\star}}{\zeta_\ell\Delta^2_{\ell,i^\star}} \nonumber \\
\label{eq:regret_bound_risk_aware_greedy_UCB}
& +\left( 1+\frac{\pi^2}{2} \right) \sum_{\ell\in\mcB\setminus\mcS^\star} \sum_{i^\star\in\mcS^\star} (\zeta_\ell \Delta_{\ell,i^\star}+(1-\zeta_\ell)\tilde{\Delta}_{\ell,i^\star}),
\end{align}
where $\delta=(\sqrt{5}-1)/2$. 
\end{theorem}

Theorem \ref{thm:regret_bound_risk_aware_greedy_UCB} shows a regret bound for Algorithm \ref{alg:risk_aware_greedy_UCB_algorithm}. The algorithm still has $\mcO(\log(n))$ regret but with a larger constant. 
This shows that introducing risk-awareness increases the learning time in the sense that $R_2[n]>R_1[n]$. This is because by rejecting a high-risk beam pair, the algorithm loses the chance to get information on that beam pair. The idea of Algorithm \ref{alg:risk_aware_greedy_UCB_algorithm} is to distribute the learning of these high-risk beam pairs (which has high cost) more evenly among the users by rejecting them with some probability. 
In other words, Algorithm~\ref{alg:risk_aware_greedy_UCB_algorithm} tradeoffs the learning speed to balance the risk of severe misalignment endured by each user at different stages of the learning. 

\section{Online beam pair refinement} \label{sec:online_beam_refinement}
In this section, we describe our beam refinement solution, which is the second layer of the two-layer online learning algorithm. 
We start with the problem formulation and then describe our modified HOO solution. 

\subsection{Problem statement} \label{sec:prob_stat_beam_refinement}

We formulate our beam pair refinement as a stochastic CAB problem. The beams are generated by progressive phase-shift and are defined by their main beam directions. The goal is to find the pointing direction of a beam pair to maximize the average channel strength of that beam pair in an online setting. Specifically, denoting $\phi^\rmt_i,\theta^\rmt_i,\phi^\rmr_i,$ and $\theta^\rmr_i$ the transmit and receive main beam directions in the azimuth and elevation of the beam pair $i$ defined in the codebook, and $\Phi^\rmt_i,\Theta^\rmt_i,\Phi^\rmr_i,$ and $\Theta^\rmr_i$ the corresponding 3dB beamwidths, the problem of refining the beam pair~$i$ can be written~as
\begin{align}
\label{eq:beam_refinement_prob}
\begin{array}{rl}
\underset{\phi^\rmt,\theta^\rmt,\phi^\rmr,\theta^\rmr}{\mathrm{maximize}} & \bbE[\gamma_i(\phi^\rmt,\theta^\rmt,\phi^\rmr,\theta^\rmr)] \\
\mathrm{subject\,\, to} & \phi^\rmt \in [\phi^\rmt_i-\Phi^\rmt_i/2,\phi^\rmt_i+\Phi^\rmt_i/2] , \\
& \phi^\rmr \in [\phi^\rmr_i-\Phi^\rmr_i/2,\phi^\rmr_i+\Phi^\rmr_i/2] , \\
& \theta^\rmt \in [\theta^\rmt_i-\Theta^\rmt_i/2,\theta^\rmt_i+\Theta^\rmt_i/2] ,\\
& \theta^\rmr \in [\theta^\rmr_i-\Theta^\rmr_i/2,\theta^\rmr_i+\Theta^\rmr_i/2] .
\end{array}
\end{align}
Any pointing direction $(\phi^\rmt,\theta^\rmt,\phi^\rmr,\theta^\rmr)$ satisfying the constraints is an arm in this problem. The space to search for the best arm is the hyperrectangle defined by the constraints, which is a continuous space. 
This means the directions are fine-tuned within the 3dB beamwidths of the original beam pair $i$ defined by the pointing direction $(\phi^\rmt_i,\theta^\rmt_i,\phi^\rmr_i,\theta^\rmr_i)$. The coarse search to within the 3dB beamwidth is supposed to be done by the beam pair selection algorithm. 

\subsection{Modified HOO for beam pair refinement} \label{sec:modified_HOO}

HOO is a CAB algorithm that runs on a tree. We start by describing the search tree. Then, we describe the flow of HOO. Finally, we explain the modifications made to the original algorithm to fit the beam refinement task. We describe the algorithm for refining a beam pair $i$. Since all the description is in the context of this beam pair $i$, we drop explicit references to beam pair $i$ here to avoid notational clutter.

We now define the search tree $\mcT$ which HOO runs on. 
Each node in the tree is a pair of transmit and receive pointing directions $(\phi^\rmt,\theta^\rmt,\phi^\rmr,\theta^\rmr)$ satisfying the constraints in \eqref{eq:beam_refinement_prob}. 
The root of the tree is the original pointing direction of the beam pair $i$ $(\phi^\rmt_i,\theta^\rmt_i,\phi^\rmr_i,\theta^\rmr_i)$ defined in the codebook. 
Each node in the tree at depth $\ell<\ell_{\max}$ has 16 children which correspond to all possible combinations of transmit and receive beam directions perturbed by $1/2^\ell$ of the beamwidths in the four variables. Denote $(\phi^\rmt_{\ell,k},\theta^\rmt_{\ell,k},\phi^\rmr_{\ell,k},\theta^\rmr_{\ell,k})$ the parameters of the $k$-th node at depth $\ell$ in $\mcT$, its set of 16 children nodes can be written using a Cartesian product as 
\begin{align}
\label{eq:set_children_nodes}
\left\{
\begin{Bmatrix}
[\phi^\rmt_{\ell,k} + \Phi_i^\rmt/2^\ell,\,\,\, \theta^\rmt_{\ell,k}]^\rmT, \\
[\phi^\rmt_{\ell,k} - \Phi_i^\rmt/2^\ell,\,\,\, \theta^\rmt_{\ell,k}]^\rmT, \\
[\phi^\rmt_{\ell,k}, \,\,\, \theta^\rmt_{\ell,k}+\Theta_i^\rmt/2^\ell]^\rmT, \\
[\phi^\rmt_{\ell,k}, \,\,\, \theta^\rmt_{\ell,k}-\Theta_i^\rmt/2^\ell]^\rmT 
\end{Bmatrix} \! \times \! 
\begin{Bmatrix}
[\phi^\rmr_{\ell,k} + \Phi_i^\rmr/2^\ell,\,\,\, \theta^\rmr_{\ell,k}]^\rmT, \\
[\phi^\rmr_{\ell,k} - \Phi_i^\rmr/2^\ell,\,\,\, \theta^\rmr_{\ell,k}]^\rmT, \\
[\phi^\rmr_{\ell,k},\,\,\, \theta^\rmr_{\ell,k}+\Theta_i^\rmr/2^\ell]^\rmT, \\
[\phi^\rmr_{\ell,k},\,\,\, \theta^\rmr_{\ell,k}-\Theta_i^\rmr/2^\ell]^\rmT 
\end{Bmatrix}  
\right\}. 
\end{align}
Using this node expansion rule, a node at depth $\ell+1$ deviates from its parent node in the pointing direction by $\mathrm{beamwidth}/2^\ell$ and depends only on $\ell$. Each depth in the tree can be thought of as a grid partitioning the search space defined by the constraints in \eqref{eq:beam_refinement_prob}. The grid becomes finer deeper in the tree (i.e., as $\ell$ increases). 

{\renewcommand{\baselinestretch}{1.0} 
\begin{algorithm}
 \caption{Modified HOO for Beam Pair Refinement}
 \label{alg:HOO_beam_refinement_algorithm}
 \begin{algorithmic}[1]
  \STATE // Initialization 
  \STATE $\mcT \leftarrow \{(0,1)\}\cup \mcC_{1,1}$ 
  \STATE $(B_{2,j},T_{2,j},\hat{\mu}_{2,j},\mathsf{Sq}_{2,j})\leftarrow (\infty,0,0,0)$ for $\forall j\in \mcC_{1,1}$
  \FOR {$n=1,2,\dots$} 
  	\STATE // Select a node in the tree to sample
  	\STATE $(\ell,k) \leftarrow (1,1)$ \COMMENT{Start from the root node} \label{Alg3:Select_node_start}
  	\STATE $\mcP\leftarrow \{ (\ell,k) \}$ 
  	\FOR { $\ell=1,\dots,\min\{\mathrm{depth}(\mcT),\ell_{\max}-1\}$ }
  		\STATE $k^\star \leftarrow \underset{j\in \mcC_{\ell,k}}{\arg\max}\, B_{\ell+1,j}$ 
  		\STATE $(\ell,k) \leftarrow (\ell+1,k^\star)$ 
  		\STATE $\mcP\leftarrow \mcP \cup \{ (\ell,k) \}$
  	\ENDFOR 
  	\STATE $(\ell_\rms,k_\rms)\leftarrow (\ell,k)$ \label{Alg3:Select_node_end}
  	\STATE Obtain the beam measurement for node $(\ell_\rms,k_\rms)$ denoted by $\gamma$ \label{Alg3:Beam_measurement}
  	\STATE // Update the learning parameters
  	\FOR [Update sample averages]{$(\ell,k)\in \mcP$} 
  		\STATE $T_{\ell,k} \leftarrow T_{\ell,k} + 1$
  		\STATE $\hat{\mu}_{\ell,k} \leftarrow (1-\frac{1}{T_{\ell,k}})\hat{\mu}_{\ell,k} + \gamma/T_{\ell,k}$
  		\STATE $\mathsf{Sq}_{\ell,k}\leftarrow \mathsf{Sq}_{\ell,k} + \gamma^2$ 
  		\STATE $\sigma_{\ell,k}^2\leftarrow ( \mathsf{Sq}_{\ell,k}- \hat{\mu}_{\ell,k}^2 T_{\ell,k}  )/T_{\ell,k}$ 
  	\ENDFOR
  	\FORALL [Update U-values]{ $(\ell,k)\in \mcT$ }
  		\STATE $U_{\ell,k}\leftarrow \left( \hat{\mu}_{\ell,k} + \sqrt{16 \hat{\sigma}_{\ell,k}^2\frac{\log(n)}{T_{\ell,k}} } \right) \nu(\ell) $ \label{Alg3:update_u_vals}
  		\STATE $U_{\ell,k}\leftarrow \infty$ if $T_{\ell,k} < \lceil \alpha_\mathrm{norm} \log(n) \rceil$ or $T_{\ell,k} < K_{\min}$ \COMMENT{Forced exploration} \label{Alg3:enforced_norm_UCB} 
  	\ENDFOR 
	\STATE // Expand a node if conditions are met 
  	\IF { $\ell_\rms < \ell_{\max} \land T_{\ell_\rms,k_\rms}> K_{\mathrm{exd}} \land (\ell_\rms,k_\rms)$ is a leaf}  
  		\STATE $\mcT \leftarrow \mcT \cup \mcC_{\ell_\rms,k_\rms}$
  		\STATE $(B_{\ell_\rms+1,j},T_{\ell_\rms+1,j},\hat{\mu}_{\ell_\rms+1,j},\mathsf{Sq}_{\ell_\rms+1,j})\leftarrow (\infty,0,0,0)$ for $\forall j\in \mcC_{\ell_\rms,j}$
  	\ENDIF 
	\STATE // Update B-values 
  	\FOR {$\ell=\ell_\rms,\ell_\rms-1,\dots,2$ } 
  		\STATE $B_{\ell,k}\leftarrow \min\{ U_{\ell,k},\max_{j\in \mcC_{\ell,k}} B_{\ell+1,j} \}, \forall $ nodes at depth $\ell$ in $\mcT$ \label{Alg3:update_B_vals}
  	\ENDFOR
  	
  \ENDFOR
 \end{algorithmic} 
\end{algorithm}
}

We now describe how the modified HOO works. A pseudo-code is shown in Algorithm \ref{alg:HOO_beam_refinement_algorithm}. It runs on a finite tree with a maximum depth of $\ell_{\max}$.  
The nodes in the tree are activated on the fly, and only the root node and its children are active at time $n=0$. Thus, the initial tree is $\mcT=\{(1,1)\}\cup \mcC_{1,1}$, where $\mcC_{\ell,k}$ denotes the set of the indices of the children of the node $(\ell,k)$.  
In each iteration, there are three main parts. First, a node is selected by traversing the active tree starting from the root following the path through nodes that have the largest B-values (line \ref{Alg3:Select_node_start}-\ref{Alg3:Select_node_end}), which is the best optimistic estimate of the average rewards of the nodes. 
The second part is the beam measurement for the selected node (line \ref{Alg3:Beam_measurement}). Lastly, after obtaining the measurement, the learning parameters are updated. If the condition is met, a node in the tree is expanded, i.e., activating its 16 children nodes. 
Note that to lower the risk of expanding a suboptimal node, it is enforced that a node can be expanded only after it is sampled $K_\mathrm{exd}$ times. 
The last part of the parameter update is the B-values. They are computed by back calculation from the sampled node back to the root (line \ref{Alg3:update_B_vals}). 
The B-value of node $(\ell,k)$ is the minimum between its own U-value and the maximum B-value of its children nodes, i.e.,
\begin{align}
B_{\ell,k}\leftarrow \min \left\{ U_{\ell,k},\max_{j\in \mcC_{\ell,k}} B_{\ell+1,j} \right\}.
\end{align} 
The U-value is similar to the UCB value, but it also accounts for the smoothness property (line \ref{Alg3:update_u_vals}).
The U-value provides an optimistic estimate of its average reward using the parameter of the node, and the maximum B-value among its children nodes provides another optimistic estimate of its reward. By taking the minimum between the two, the obtained B-value provides a refined optimistic estimate of the average reward of the node.

We introduce three main modifications to the original HOO tailored  to the beam refinement setting. 
The first one is the use of a finite tree. The original HOO assumes an infinite tree to represent the arm space. Since small adjustments (e.g., 1/8 of the beamwidth) have a small impact on the gain, a finite tree of maximum depth $\ell_{\max}$ is used instead to save computation and storage for the learning parameters. The second one is the smoothness bound. The original HOO assumes an additive offset. Due to the multiplicative nature of the antenna gains, a multiplicative smoothness coefficient $\nu(\ell)$ as shown in line~\ref{Alg3:update_u_vals} is more suitable. The coefficient is computed using Lemma~\ref{lem:smoothness_coefficient} which will be detailed in the last part of this subsection. 

The third modification is the confidence margin. 
Because the original margin $\sqrt{2\log(n)/T_{\ell,k}}$ is too loose in our setting, we propose to use that of the norm-UCB (line \ref{Alg3:update_u_vals}) \cite{Auer2002}. 
The margin $\sqrt{2\log(n)/T_{\ell,k}}$ is derived from the Chernoff-Hoeffding inequality, which is applicable to any distribution with the support in $[0,1]$. While normalizing the channel strengths by a large enough number will approximately guarantee that the support is within $[0,1]$, the average typically takes a value much less than 1 and the margin $\sqrt{2\log(n)/T_{\ell,k}}$ is too loose for reasonable learning horizons. 
The reason that the average is much smaller than 1 is the small scale fading nature of the wireless channel. Fading is due to the multipath effect and can cause the maximum instantaneous channel strength to be much larger than the average \cite{wirelessLabTextbook}. A good property of the new margin is that the sample variance is also used. 
Note that to enable regret analysis, the norm-UCB algorithm requires each arm be sampled at least $\lceil\alpha_\mathrm{norm}\log(n)\rceil$ at time $n$ with $\alpha_\mathrm{norm}=8$ \cite{Auer2002}. This is enforced by setting the U-values of the nodes that need to be explored to infinity (see line \ref{Alg3:enforced_norm_UCB}). Note that we also introduce the condition $T_{\ell,k}<K_{\min}$, which is used to ensure that there are at least $K_{\min}$ samples of the node for computing the sample variance. This is needed when using a small $\alpha_{\mathrm{norm}}$. 


We next state a lemma defining the smoothness property of the objective function in \eqref{eq:beam_refinement_prob}. The lemma is useful for computing the smoothness coefficient $\nu(\ell)$. 
\begin{lemma} \label{lem:smoothness_coefficient}
Assume a single-path {\it azimuth} PAS with the optimal beam direction $\phi^\star$ with isotropic transmit antenna, $G(\cdot;\phi_0)$ the normalized gain of the beam pattern pointing at $\phi_0$ assumed to be decreasing and concave in $[\phi_0,\phi_0+\Phi/2]$ with $\Phi$ denoting the 3 dB beamwidth (e.g., true for a uniform planar array), for a receive pointing direction $\phi_0$ such that $|\phi^\star-\phi_0|\le\Delta\phi\le \Phi/2$, 
\begin{align}
\bar{\gamma}(\phi_0) / G(\phi_0+\Delta\phi;\phi_0) \ge \bar{\gamma}(\phi^\star). 
\end{align}
Moreover, for a general PAS with the support within $[\phi_0-\Psi,\phi_0+\Psi]$ with $\Psi\le \Phi/2$, 
\begin{align}
\bar{\gamma}(\phi_0)/G(\phi_0+\Delta\phi;\phi_0) \ge \bar{\gamma}(\phi^\star) - \mathsf{Err},
\end{align}
where $\mathsf{Err}\ge0$ is a residual term that depends on the shape of the PAS and 
$\mathsf{Err}\to0$ as $\Delta\phi\to 0$. 
\end{lemma}

We now explain how to determine the smoothness coefficient $\nu(\ell)$ based on Lemma~\ref{lem:smoothness_coefficient}. While we state Lemma \ref{lem:smoothness_coefficient} assuming an isotropic transmit antenna to avoid tedious notations, the same argument applies when we also include the transmit beam pattern. In particular, denoting $\phi^{\rmr\star},\phi^{\rmt\star}$ the optimal receive and transmit beam direction, 
\begin{align}
\label{eq:azimuth_smoothness_bound_txrx}
\frac{\bar{\gamma}(\phi_0^{\rmr},\phi_0^{\rmt})}{G_\rmr(\phi_0^\rmr+\Delta\phi^\rmr;\phi_0^\rmr)G_\rmt(\phi_0^{\rmt}+\Delta\phi^\rmt;\phi_0^{\rmt})} \ge \bar{\gamma}(\phi^{\rmr\star},\phi^{\rmt\star}) - \mathsf{Err}.
\end{align}
When steering the elevation only, we get the same relation as \eqref{eq:azimuth_smoothness_bound_txrx}. If we assume square UPAs, the beam pattern in the azimuth and elevation will be the same. Since we only change the azimuth or elevation but not both per \eqref{eq:set_children_nodes}, the smoothness coefficient is given by
\begin{align}
\nu(\ell) & = a/(G_\rmr(\phi_\ell^\rmr+\Phi/2^\ell;\phi_\ell^\rmr)G_\rmt(\phi_\ell^{\rmt}+\Phi/2^\ell;\phi_\ell^{\rmt}))  \\
& \simeq a/g^2(\mathrm{beamwidth}/2^\ell),
\end{align}
where $a>1$ is a correction coefficient to account for $\mathsf{Err}$ if deemed necessary. For convenience, we approximate the gain by $g(\cdot)$ the beam pattern at broadside as a function of the deviation from the broadside direction. Note that we will need $a$ only for large $\Delta\phi$. Deeper in the tree, the change in the angle is small and thus $\mathsf{Err}$ will become negligible. Also, for the sake of clear argument, we restrict $\Psi\le\Phi/2$, but with a more elaborate choice of the coefficient of $\mathsf{Err}$ in the proof, we can allow $\Psi$ to be larger. This, however, is not a big concern in our setting because $\Delta\phi$ will be $\Phi/4$ or less and $\mathsf{Err}$ is restricted to a small value already.


\section{Numerical results} \label{sec:numerical_results}

We start with the general setting of our numerical evaluations.
As described in Section \ref{sec:data_model}, our codebook for $16\times 16$ UPA has 271 beams and thus there are $271^2$ beam pairs. 
Using the heuristic screening to get $\hat{\mcB}$ as explained in Section \ref{sec:prob_stat_beam_selection} with the initial database size of $N=5$ and $C=200$, the size of the set of beam pairs to be learned $|\hat{\mcB}|$ is typically between 400 and 600 depending on the simulation run. As mentioned in Section \ref{sec:data_model}, we generated 10,000 channel samples using ray-tracing. To eliminate the effect of the ordering of the channel samples on the learning performance, the evaluation metrics are averaged over 100 simulation runs, where in each run we randomly permute these 10,000 channel samples. We apply a moving average with a window size of 50 time steps to better show the trends.

As an evaluation metric, we use the 3 dB power loss probability and the gain defined as the inverse of the power loss in \eqref{eq:powLoss_def}. The 3 dB power loss probability (i.e., $c=2$ in \eqref{eq:powLossProb_def}) measures how often the selected beam pair has a loss larger than 3 dB as compared to the best beam pair selected by exhaustive search, and thus capturing the beam alignment accuracy. This metric, however, is not suitable for evaluating the beam pair refinement because it cannot capture the improvement over the exhaustive search in the original codebook. 
Allowing the refinement, a beam pair better than the best in the original codebook can be selected resulting in power loss taking a value less than one, or equivalently, a positive gain in dB.

The rest of the section is divided into three parts. Section \ref{sec:numRes_beam_sel} evaluates the beam pair selection alone without the refinement option. 
Section \ref{sec:numRes_beam_refinement} assesses the performance of the beam pair refinement assuming an offline learning for the beam pair selection. Section \ref{sec:numRes_integrated_sol} provides evaluations of the integrated solution 
incorporating both components. 

\subsection{Online Beam Pair Selection} \label{sec:numRes_beam_sel}

\begin{figure}
	\centering
	\includegraphics[width=0.9\columnwidth]{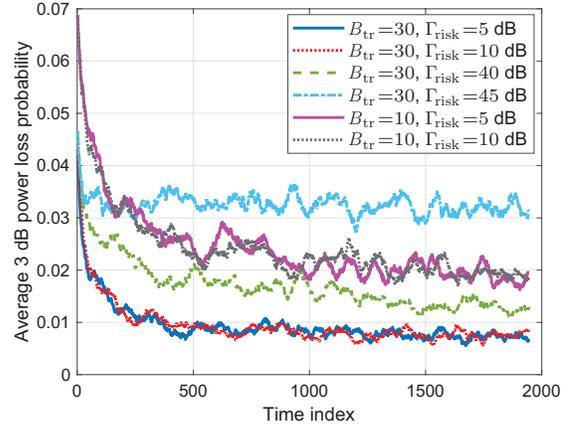}
	\caption{Average 3dB power loss probability using the proposed risk-aware greedy UCB algorithm with different training budgets $\Btr$ and risk thresholds $\Grisk$. For both $\Btr=10$ and 30, the plots show similar learning behavior. A smaller training budget $\Btr=10$ provides less accuracy beam alignment. The plots using different $\Grisk$ show that the performance is not sensitive to $\Grisk$ as long as it is not too large.}
	\label{fig:Avg3dBPowLossProb_vs_time_diff_Nb_diff_A}
\end{figure}

This subsection evaluates the performance of the proposed risk-aware greedy UCB algorithm without the beam refinement option. There are two parameters to be decided when running Algorithm \ref{alg:risk_aware_greedy_UCB_algorithm}: the training budget $\Btr$ and the risk threshold $\Grisk$ in \eqref{eq:risk_signal}. We note that our solution does not require that $\Btr$ be fixed, but for simplicity, we assume that the same $\Btr$ is used during the entire learning horizon. Fig. \ref{fig:Avg3dBPowLossProb_vs_time_diff_Nb_diff_A} shows the average 3dB power loss probability versus time for $\Btr=10$ and 30 with different $\Grisk$. 
We can confirm from the figure that using a larger training budget $\Btr$ leads to lower 3dB power loss probability, i.e., more accurate beam alignment. The learning seems to have two phases: the fast improvement phase in the early time steps and the slower improvement phase after that. 
For $\Btr=30$ and $\Grisk=5$dB, this phase change happens at around time index 500. 
The slower learning phase starts when the algorithm has identified high-risk beam pairs (with some certainty) and learns those beam pairs at a slow pace due to the rejection mechanism. Regarding the risk threshold, the results show that the algorithm is not sensitive to the choice of $\Grisk$. As long as $\Grisk$ is not too large (e.g., less than 40dB), it performs well. 
The main reason for this behavior is due to the effect of the replacement selection method (line \ref{Alg2:alternative_sel_start}-\ref{Alg2:alternative_sel_end} in Algorithm \ref{alg:risk_aware_greedy_UCB_algorithm}) that selects beam pairs to replace those rejected; even if a good beam pair gets rejected due to risk overestimation (when using a small $\Grisk$), it will likely be picked up by the replacement selection.


\begin{figure}
	\centering
	\includegraphics[width=0.9\columnwidth]{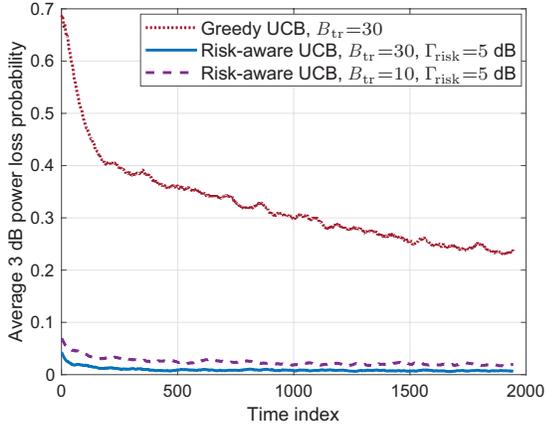}
	\caption{Performance comparison of greedy UCB with and without risk-awareness. The performance is an order of magnitude worse without risk-awareness. 
	This is because the risk-aware greedy UCB uses the risk estimates to control the number of high-risk beam pairs selected in the subset $\mcS$ reducing the probability of severe misalignment.
	}
	\label{fig:baseline_comparison_for_online_beam_selection}
\end{figure}

Fig. \ref{fig:baseline_comparison_for_online_beam_selection} shows a performance comparison of the greedy UCB algorithm with and without the risk-awareness. 
The performance without risk-awareness is an order of magnitude worse than that with risk-aware. This might seem a bit counterintuitive because the regret bound of the risk-aware algorithm is higher. One way to understand this behavior is this. The goal of the UCB selection is to reach a state where we can ensure that a suboptimal arm is {\it not} selected with high probability (call this the optimal state). To reach the optimal state, each arm has to be sampled enough times ($T_{\ell,i^\star}^{(0)}$ in the derivation in Appendix \ref{Apd:proof_UCB_greedy}). Algorithm \ref{alg:greedy_UCB_algorithm} samples the arms to reach this state fast, but it will expose early stage users to select more high-risk arms. Algorithm~\ref{alg:risk_aware_greedy_UCB_algorithm} balances the number of high-risk arms at any round by the rejection mechanism which results in a slower speed to reach the optimal state, i.e., a slower learning speed. By not exposing a user to too many high-risk arms, Algorithm \ref{alg:risk_aware_greedy_UCB_algorithm} can ensure that the regret each user has to endure is not too large.  In other words, although the cumulative regret is smaller (at a large enough time), users in early stages of Algorithm \ref{alg:greedy_UCB_algorithm} have to sacrifice. 
Algorithm \ref{alg:risk_aware_greedy_UCB_algorithm} distributes the regret more evenly among the users at different learning stages. We note that because of the large number of arms (400 to 600 as noted earlier), the time to reach the optimal state is large and Algorithm~\ref{alg:greedy_UCB_algorithm} is not practical as an online solution as shown in Fig. \ref{fig:baseline_comparison_for_online_beam_selection}.  

\begin{figure}
	\centering
	\includegraphics[width=0.9\columnwidth]{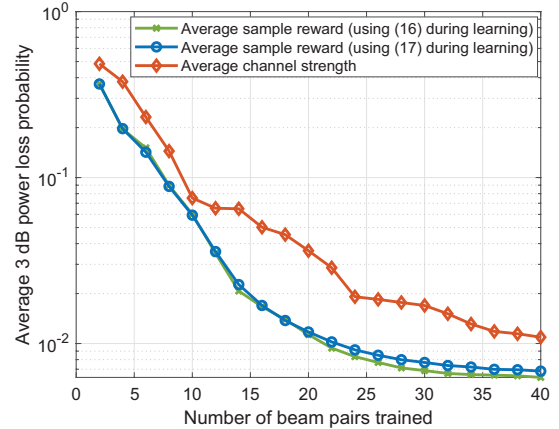}
	\caption{A comparison of the accuracy of the selection set produced by the average of the proposed reward signal ($\hat{P}_\rmopt(i)$) and the more intuitive choice of average channel strength. The performance when using $\hat{P}_\rmopt(i)$ is consistently better for all training budgets. The comparison when using the proposed practical reward signal \eqref{eq:alternative_reward_def} as opposed to the ideal reward signal \eqref{eq:true_reward_def} shows negligible performance loss. }
	\label{fig:afterlearning_accuracy_of_selection_metrics}
\end{figure}

The last part of this subsection shows the effectiveness of our choice of the reward signal in \eqref{eq:alternative_reward_def}. Specifically, we compare the accuracy of the beam selection using the average sample rewards ($\hat{P}_\rmopt(i)$) versus the more intuitive choice of average channel strength $\bar{\gamma}_i$. We also compare it with the case where we assume that the ideal reward defined in \eqref{eq:true_reward_def} is available to the algorithm during the learning. To evaluate this, we let the online learning run for 2000 time steps. 
We, then, use $\hat{P}_\rmopt(i)$ and $\bar{\gamma}_i$ estimated at time step 2000 to get two sets of beam selections and evaluate the two sets over 500 channel samples. We use $\Btr=30$ and $\Grisk=5$dB for the online learning. Fig.~\ref{fig:afterlearning_accuracy_of_selection_metrics} shows the 3dB power loss probability against the number of beam pairs trained.  
We can see that the beam pair selection using $\hat{P}_\rmopt(i)$ is more accurate than using the average channel strengths. Also, the plots show that the degradation due to the use of the proposed alternative and practical reward signal in \eqref{eq:alternative_reward_def} during the learning results in negligible loss. 
These results confirm the effectiveness of our choice of the reward signal in~\eqref{eq:alternative_reward_def}. 

\redtext{


Before moving on to the beam refinement part, we provide some comments regarding the overhead and the requirement on the position accuracy. Regarding the accuracy of the alignment, as reported in \cite{Inverse-fingerprint-preprint}, there is negligible loss in the data rate as compared to the exhaustive search when the average 3dB power loss probability is less than 1\%. We can see in Fig.~\ref{fig:Avg3dBPowLossProb_vs_time_diff_Nb_diff_A} that the average 3dB power loss falls below 2\% quickly and reaches 1\% within about 300 time steps with $\Btr=30$ and $\Grisk=5$dB. 
The overhead per beam alignment attempt, which affects the instantaneous performance, is the same as the offline method, and the detailed study has been reported in our prior work \cite{Inverse-fingerprint-preprint}. We thus only provide a summary of the main results here for completeness. 

The overhead per beam alignment attempt of the proposed method is determined by the training budget $\Btr$. When using $16\times16$ arrays at both the transmitter and receiver, $\Btr=30$ was shown in \cite{Inverse-fingerprint-preprint} to provide negligible performance loss compared to the exhaustive search. This is the reason we chose $\Btr=30$ to run our online beam pair selection algorithm so that the overhead in each beam training attempt is kept the same as in the offline approach from \cite{Inverse-fingerprint-preprint}. In \cite{Inverse-fingerprint-preprint}, we compared the performance of the offline approach with two existing methods: IEEE 802.11ad and the method that uses the position only without the past beam measurements. The training overhead of our approach is less than a few percents of that of the IEEE 802.11ad method (the larger the array the smaller the percentage). Detailed comparison in the mobility context shows that 
our proposed beam alignment can support large arrays at high vehicular speed, while the IEEE 802.11ad struggles and its beam training time can eat up all the communication time before realignment is needed.
It was also shown that using only the position results in severe performance loss when the blockage probability is not negligible (e.g., in a dense traffic). In other words, the value of past beam measurements increases with the LOS blockage probability. 

Regarding the position accuracy, our method only requires that the position is accurate enough to identify the location bin. To investigate the sensitivity to position error, in \cite{Inverse-fingerprint-preprint} we evaluated the performance for bin sizes ranging from 2m to 5m. The results show that for $16\times16$ arrays, there was no performance difference. Thus, we use the 5m location bin in our evaluation in this paper. When using a larger array such as $32\times32$, the position accuracy requirement increases and the results show that 
a smaller bin size such as 2m yields better performance. 
While the results show that larger arrays require more accurate position information, even for the large $16\times 16$ array considered in this paper, only a few meters of accuracy is required.
}

\subsection{Online beam pair refinement} \label{sec:numRes_beam_refinement}

To evaluate the performance of the beam refinement on its own, we perform an offline beam pair selection using the MinMisProb method from \cite{Inverse-fingerprint-preprint} before running the beam refinement. In each simulation run, we use the first 300 channel samples to determine the selection set $\mcS$, and then we run the beam pair refinement on each of the beam pairs in $\mcS$ where we set the training budget to $\Btr=30$. 
For a baseline comparison, we implement an MAB solution using the norm-UCB algorithm from \cite{Auer2002}. The MAB solution is run on the leaves of the search tree, and thus the number of arms is $16^{\ell_{\max}-1}$. 
Besides $\Btr$, we also need to specify the maximum tree depth $\ell_{\max}$ and the forced exploration parameter $\alpha_\mathrm{norm}$. We use $K_{\min}=3$ and $K_\mathrm{exd}=10$.

\begin{figure}
	\centering
	\subfloat[HOO vs. MAB with $\ell_{\max}=3$. ]{\includegraphics[width=0.9\columnwidth]{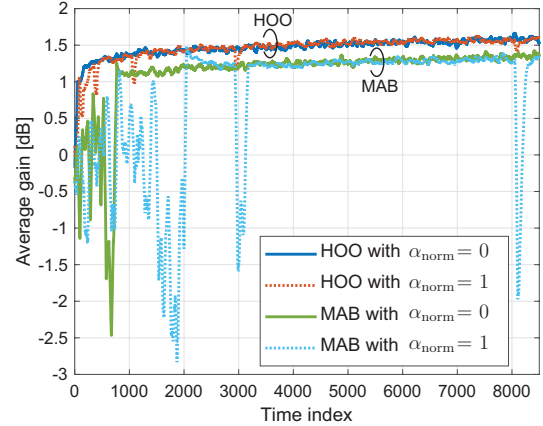}
	\label{fig:comparison_HOO_MAB_diff_alphaNorm}} \\
	\subfloat[HOO with different $\ell_{\max}$ and $\alpha_{\mathrm{norm}}=0$. ]{\includegraphics[width=0.9\columnwidth]{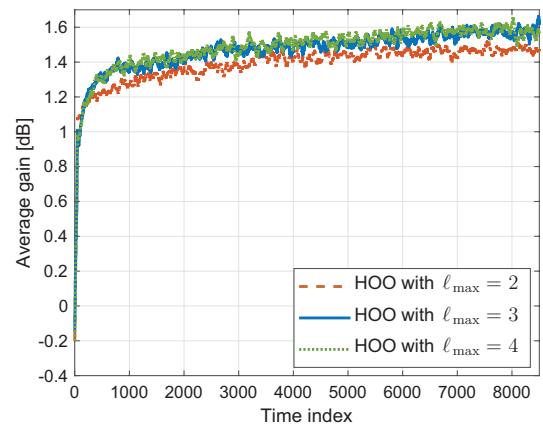}
		\label{fig:comparison_HOO_diff_lmax}}
	\caption{A comparison of HOO and MAB with different $\alpha_{\mathrm{norm}}$ and $\ell_{\max}$. Fig. (a) compares the performance when $\ell_{\max}=3$. MAB does not use the hierarchical structure of the search tree as HOO and suffers a larger exploration penalty. The penalty is even more severe as $\alpha_{\mathrm{norm}}$ increases. The results show that the forced exploration is not needed and $\alpha_{\mathrm{norm}}=0$ should be used. Fig. (b) compares the performance of HOO when using different $\ell_{\max}$. There is negligible gain for setting $\ell_{\max}$ beyond 3. We also see that HOO does not have extra degradation due to exploration when we increase $\ell_{\max}$. }
	\label{fig:comparison_HOO_MAB}
\end{figure}

We start by comparing the performance of MAB and our modified HOO solution in Fig. \ref{fig:comparison_HOO_MAB_diff_alphaNorm}. We can see that HOO learns much faster by leveraging the tree structure. We can see the cost of exploration of MAB in the initial stage, where each arm has to be tried $K_{\min}$ times. 
Using the search tree, starting from the root, HOO will first explore the nodes at depth 2. At depth 3, it explores only the children nodes of promising nodes at depth 2, and this goes on until reaching $\ell_{\max}$. This way, HOO does not have to sample all the leaves uniformly to explore the whole arm space leading to more efficient exploration than MAB.

We next show the effect of  $\alpha_\mathrm{norm}$ and $\ell_{\max}$ on the performance. 
We noted earlier that $\alpha_\mathrm{norm}=8$ is required to derive a regret bound in \cite{Auer2002}. Forcing exploration this way with $\alpha_\mathrm{norm}$ turns out to result in bad performance for our applications as shown in Fig. \ref{fig:comparison_HOO_MAB_diff_alphaNorm}. The dips in the gains are due to this forced exploration, and the intervals between dips decrease as $\alpha_\mathrm{norm}$ increases. Note that even with $\alpha_\mathrm{norm}=0$, both MAB and HOO still explore because of the confidence margin of the norm-UCB $\sqrt{16\hat{\sigma}^2_{\ell,k}\log(n)/T_{\ell,k}}$. 
Fig. \ref{fig:comparison_HOO_MAB_diff_alphaNorm} shows that for both MAB and HOO, $\alpha_\mathrm{norm}=0$ provides the best performance. Fig. \ref{fig:comparison_HOO_diff_lmax} compares the performance of HOO for $\ell_{\max}=2,3$ and $4$ with $\alpha_{\mathrm{norm}}=0$. We can see that a larger $\ell_{\max}$ improves the gains, which is expected since it allows a more refined search. Remarkably, thanks to the structure of the search tree, a larger $\ell_{\max}$ does not require more cost in the exploration. Since the performance improvement is quite small and the number of nodes in the tree increases quickly, we use $\ell_{\max}=3$ from now on.



\begin{figure}
	\centering
	\includegraphics[width=0.9\columnwidth]{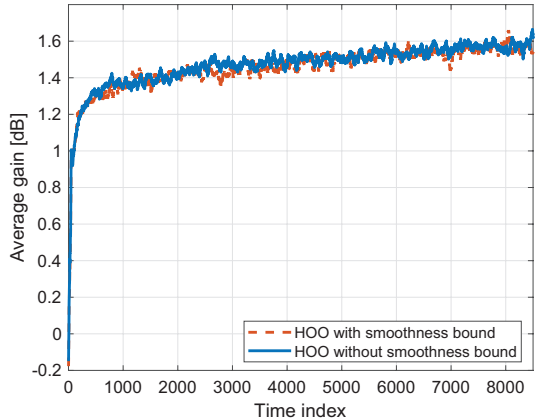}
	\caption{A comparison of HOO with and without the smoothness coefficient $\nu(\ell)$ (for computing the U-values). The smoothness coefficient shows negligible effect. This is likely because the refinement problem searches locally within the 3dB beamwidth. Since this is in the vicinity of the optimal point, it is not possible to eliminate search regions using the smoothness bound.}
	\label{fig:comparison_with_without_smoothness_bound}
\end{figure}

Fig. \ref{fig:comparison_with_without_smoothness_bound} compares the HOO beam refinement with and without the smoothness coefficient $\nu(\ell)$. The performance difference is negligible. This is likely because the search region in our problem is already confined to a small local region (within the 3dB beamwidths of the selected beam pair) so that the constraint derived from the smoothness property does not have much value. This has a welcoming implication. The algorithm can be expected to be robust to small irregularity in the detailed shape of the beam patterns (thus affecting the exact smoothness property), which can be expected with real hardware. 

\subsection{Integrated online learning solution} \label{sec:numRes_integrated_sol}
This subsection evaluates the performance when combining the beam pair selection and refinement together. One thing that needs to be specified when combining the two is when to start the refinement for a selected beam pair. We consider the following three variations to start the beam refinement:
\begin{enumerate}
	\item Refine all: The beam pair refinement is started for any beam pair from the first time it is selected by the online beam pair selection algorithm. This is the most straightforward way to combine the two components. 
	\item Refine after $X_\mathrm{tot}[i]>0$: The refinement of the beam pair $i$ starts from the time step that the beam pair $i$ receives a reward, i.e., when $X_\mathrm{tot}[i]$ becomes positive. The point for this option is that the algorithm only refines those beam pairs deemed to be most promising. 
	\item Refine after $n_0$ time steps: The beam refinement of all selected beam pairs starts after running the online beam pair selection for $n_0$ time steps. 
	The rationale for this option is to prevent the beam pair refinement algorithm from affecting the learning of the beam pair selection algorithm.
	This option allows the beam pair selection to run for a while so that it stabilizes to some extent before starting the beam pair refinement. 
\end{enumerate}
While it seems more efficient to focus the refinement on promising beam pairs only as in Option~2, refining suboptimal beam pairs as well will maximize their average received signal and could reduce the risk of large power loss. Thus, it is not obvious which option provides the best performance. 

\begin{figure}
	\centering
	\includegraphics[width=0.9\columnwidth]{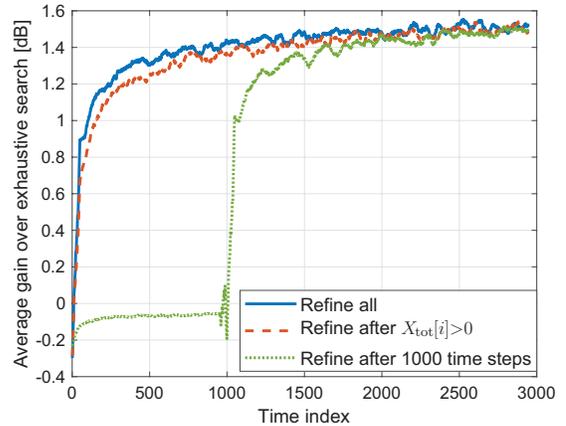}
	\caption{A comparison of average gain of the integrated solution with the three options for when to start the beam refinement. 
	The plots show no negative impact of the beam pair refinement on the online learning for beam pair selection. It is best to start the refinement simultaneously with the online beam pair selection. 
	}
	\label{fig:usingHOO_with_diff_start_refine}
\end{figure}

Fig. \ref{fig:usingHOO_with_diff_start_refine} compares the average gains over the exhaustive search (on the original codebook) of the three options. 
Here, $\Btr=30$, $\Grisk=5$dB, $\ell_{\max}=3$, $\alpha_{\mathrm{norm}}=0$, and no smoothness coefficient is used (i.e., $\nu(\ell)=1$). 
We can see that the first option, which is also the most straightforward one, provides the best performance. Focusing just on promising beam directions as in the second option performs quite well but is slightly worse than the first option. The results show that there is no benefit in waiting for some time before enabling the beam refinement as in the third option.

\section{Conclusions} \label{sec:conclusions}
In this paper, we proposed position-based online learning algorithms for beam pair selection and refinement. 
We used the MAB framework to develop a risk-aware greedy UCB algorithm for beam pair selection and a modified HOO for the beam pair refinement. Combining the two solutions together, we can gain up to about 1.5dB over the received power obtained by exhaustive search over the original beam codebook before refinement. The learning is fast and it achieves an average gain of about 1dB within the first 100 time steps. 
While we only use position in this paper, more side information from sensors on devices or the BS about the environment will help further reduce the beam training overhead. 
As shown in this work, even efficient learning algorithms can be impractical without risk-awareness because the focus is on cumulative rather than instantaneous performance. 
Therefore, we believe risk-awareness is a key to developing practical online learning solutions to take full advantage of these sensors to enable fast and efficient mmWave communications.


\appendices
\section{Proof of Theorem \ref{thm:regret_bound_greedy_UCB}} \label{Apd:proof_UCB_greedy}
The regret is non-zero when one or more suboptimal beam pairs are selected. Thus, the total expected regret can be bounded by the average number of times suboptimal pairs are selected. We note that this derivation follows the steps of the UCB1 derivation from \cite[Theorem 1]{Auer2002} with the exception of the multiple-play setting. We provide the full details for completeness and readability. 
Denote $\ell$ and $i^\star$ the indices of a suboptimal and optimal pair. Denote $T_{\ell,i^\star}[n]$ the number of times $\ell$ is selected instead of $i^\star$ up to time $n$, the expected regret is
\begin{align}
R_{\ell,i^\star}[n] = \bbE\left[ T_{\ell,i^\star}[n] \right]  \Delta_{\ell,i^\star}, 
\end{align}
where $\Delta_{\ell,i^\star}$ is the optimality gap defined in \eqref{eq:optimality_gap_def}. This is because whether $\ell$ is selected or not at time $n$ depends on the rewards up to time $n-1$ and the loss depends only on the rewards at~$n$. Thus, the two are independent by the assumption of independent reward signals across time.

We will now compute a bound for $\bbE\left[ T_{\ell,i^\star}[n] \right]$. A {\it necessary} condition for the pair $\ell$ to be selected instead of the pair $i^\star$ is that $\UCB_\ell>\UCB_{i^\star}$. After the pair $\ell$ has been selected $T_{\ell,i^\star}^{(0)}$ times, the number of times $\ell$ is selected instead of $i^\star$ up to time $n$ can be bounded by
\begin{align}
\label{eq:bound_times_selected1}
& T_{\ell,i^\star} [n] \le T_{\ell,i^\star}^{(0)} + \sum_{t=t_0}^{n}\bone\left \{\UCB_\ell \ge \UCB_{i^\star},T_\ell[t-1]\ge T_{\ell,i^\star}^{(0)} \right \} \\
& = T_{\ell,i^\star}^{(0)} + \sum_{t=t_0}^{n}\bone \left \{\hat{P}_\rmopt(\ell)+c_{t-1,T_\ell[t-1]} \ge \right . \nonumber \\ \label{eq:bound_times_selected2}
& \hspace{1.6cm} \left . \hat{P}_\rmopt(i^\star)+c_{t-1,T_{i^\star}[t-1]},T_\ell[t-1]\ge T_{\ell,i^\star}^{(0)} \right \}
\\ 
& \le T_{\ell,i^\star}^{(0)} + \sum_{t=t_0}^{n}\bone \left \{ \max_{T_{\ell,i^\star}^{(0)}<u_\ell<t}\left \{\hat{P}_\rmopt(\ell)+c_{t-1,u_{\ell}} \right \} \ge  \right. \nonumber \\ 
\label{eq:bound_times_selected3}
& \hspace{1.6cm} \left. \min_{0<u<t}\left \{\hat{P}_\rmopt(i^\star)+c_{t-1,u} \right \}  \right \} 
\\ & \label{eq:bound_times_selected4}
\le T_{\ell,i^\star}^{(0)} + \sum_{t=1}^{n-1} \sum_{u=1}^{t-1} \sum_{u_\ell=T_{\ell,i^\star}^{(0)}}^{t-1}\!\!\!\!  \bone \! \left \{ \hat{P}_\rmopt(\ell)+c_{t,u_{\ell}} \ge \hat{P}_\rmopt(i^\star)+c_{t,u} \right \}. 
\end{align}
where $t_0\ge T_{\ell,i^\star}^{(0)}$ and $c_{t,u}=\sqrt{2\log(t)/u}$ is the confidence margin. 
For $\{ \hat{P}_\rmopt(\ell)+c_{t,u_{\ell}} \ge \hat{P}_\rmopt(i^\star)+c_{t,u} \}$ to be true, at least one of the followings must hold
\begin{align}
\label{eq:clean_event_cond1}
\hat{P}_\rmopt(i^\star) & \le P_\rmopt(i^\star) - c_{t,u} \\
\label{eq:clean_event_cond2}
\hat{P}_\rmopt(\ell) & \ge P_\rmopt(\ell) + c_{t,u_\ell} \\
\label{eq:clean_event_cond3}
P_\rmopt(i^\star) & < P_\rmopt(\ell) + 2 c_{t,u_\ell}. 
\end{align}
Note that \eqref{eq:clean_event_cond1} means the UCB value underestimates the true reward of pair $i^\star$, and \eqref{eq:clean_event_cond2} means the UCB value overestimates the true reward of pair $\ell$ by larger than the corresponding confidence margins. Setting $T_{\ell,i^\star}^{(0)}=\lceil 8\log(n)/\Delta_{\ell,i^\star}^2 \rceil$, it can be shown that \eqref{eq:clean_event_cond3} is impossible \cite[p. 243]{Auer2002}, and we can bound $\bbE[T_{\ell,i^\star}[n] ]$ by
\begin{align}
\bbE[T_{\ell,i^\star}[n]] & \le \left\lceil \frac{8\log(n)}{\Delta_{\ell,i^\star}^2} \right \rceil + 
\sum_{t=1}^{n-1} \sum_{u=1}^{t-1} \sum_{u_\ell=T_{\ell,i^\star}^{(0)}}^{t-1} \bbP [\text{\eqref{eq:clean_event_cond1} is true} ] \nonumber \\
\label{eq:expected_time_suboptimal1} 
& \hspace{3cm} + \bbP [ \text{\eqref{eq:clean_event_cond2} is true} ]
\\ 
\label{eq:expected_time_suboptimal2}
& \le \left\lceil \frac{8\log(n)}{\Delta_{\ell,i^\star}^2} \right \rceil + 
\sum_{t=1}^{\infty} \sum_{u=1}^{t-1} \sum_{u_\ell=T_{\ell,i^\star}^{(0)}}^{t-1} (t^{-4} + t^{-4})
\\ \label{eq:expected_time_suboptimal3}
& \le \left\lceil \frac{8\log(n)}{\Delta_{\ell,i^\star}^2} \right \rceil + 
2 \sum_{t=1}^{\infty} \sum_{u=1}^{t} \sum_{u_\ell=1}^{t} t^{-4} 
\\ \label{eq:expected_time_suboptimal4}
& \le \left\lceil \frac{8\log(n)}{\Delta_{\ell,i^\star}^2} \right \rceil + 2\frac{\pi^2}{6}. 
\end{align}
The second line in \eqref{eq:expected_time_suboptimal2} follows because the probability terms can be shown to be bounded by $t^{-4}$ using the Chernoff-Hoeffding inequality \cite{Auer2002}. 

The total regret bound follows by summing all pairs of optimal and suboptimal beam pairs: 
\begin{align}
R_1[n] \le \sum_{\ell\in\hat{\mcB}\setminus\mcS^\star}\sum_{i^\star\in\mcS^\star} \bbE[T_{\ell,i^\star}[n]] \Delta_{\ell,i^\star}.
\end{align}
Substituting \eqref{eq:expected_time_suboptimal4} in and after some algebra, we obtain \eqref{eq:regret_bound_greedy_UCB}. 

\section{Proof of Theorem \ref{thm:regret_bound_risk_aware_greedy_UCB}} \label{Apd:proof_risk_aware_UCB_greedy}
The derivation follows similarly to that of Theorem \ref{thm:regret_bound_greedy_UCB}, but we need to be careful about the rejection mechanism. Even if a pair is selected by the greedy UCB selection, it will not be used for the training if it is rejected. Let $\tilde{T}_{\ell,i^\star}$ and $T_{\ell,i^\star}$ be the number of times the pair $\ell$ is selected instead of the pair $i^\star$ and the number of times it is accepted for beam training, respectively. We proceed similarly to obtain a bound similar to  \eqref{eq:bound_times_selected4} given~by 
\begin{align}
& \tilde{T}_{\ell,i^\star}[n] \le \tilde{T}_{\ell,i^\star}^{(0)} + \sum_{t=1}^{n-1} \sum_{u=1}^{t-1} \sum_{\tilde{u}_\ell=\tilde{T}_{\ell,i^\star}^{(0)}}^{t-1} \bone\{ \hat{P}_\rmopt(\ell)+c_{t,u_{\ell}} \ge  \nonumber \\
\label{eq:bound_times_selected}
& \hspace{5cm} \hat{P}_\rmopt(i^\star)+c_{t,u} \}.
\end{align}
To compute the bound on $\bbE\left[ \tilde{T}_{\ell,i^\star}[n] \right]$, we again use \eqref{eq:clean_event_cond1}-\eqref{eq:clean_event_cond3}. The probability bounds on \eqref{eq:clean_event_cond1} and \eqref{eq:clean_event_cond2} are still applicable.
Because of the rejection, we cannot guarantee that \eqref{eq:clean_event_cond3} is impossible, but we can bound its probability. Note that $\tilde{u}_\ell$ is the number of times the pair $\ell$ is selected, and $u_\ell$ is the number of times it is accepted for beam training. 
It can be shown that \eqref{eq:clean_event_cond3} is impossible if $u_\ell > \frac{8\log(t)}{\Delta_{\ell,i^\star}^2}$ \cite[p. 243]{Auer2002}. Thus, we can bound the probability that \eqref{eq:clean_event_cond3} holds by $\bbP\left[ u_\ell \le \frac{8\log(t)}{\Delta_{\ell,i^\star}^2} \right]$. 
With the acceptance probability $\zeta_{\ell}$, we have $\bbE[u_\ell]=\bbE[\tilde{u}_\ell]\zeta_\ell$. Setting $\tilde{T}_{\ell,i^\star}^{(0)}=\lceil 8\log(n)/(\zeta_{\ell}\delta^2\Delta_{\ell,i^\star}^2) \rceil$, we get the following bound  
\begin{align}
\label{eq:bound_prob_gap_false1}
\bbP\left[ \left. u_\ell \le \frac{8\log(t)}{\Delta_{\ell,i^\star}^2} \right| \tilde{u}_\ell=t \right] & \le \bbP\left[ \left. u_\ell \le \frac{8\log(t)}{\Delta_{\ell,i^\star}^2} \right| \tilde{u}_\ell=\tilde{T}_{\ell,i^\star}^{(0)} \right] \\ 
\label{eq:bound_prob_gap_false2}
& \hspace{-2cm} \le \bbP\left[ \left. u_\ell \le \frac{8\log(n)}{\Delta_{\ell,i^\star}^2} \right| \tilde{u}_\ell=\tilde{T}_{\ell,i^\star}^{(0)} \right] \\
\label{eq:bound_prob_gap_false3}
& \hspace{-2cm} \le n^{-4} .
\end{align}
Here, \eqref{eq:bound_prob_gap_false1} follows because $t\ge t_0 \ge \tilde{T}_{\ell,i^\star}^{(0)}$, \eqref{eq:bound_prob_gap_false2} holds because $t < n$, and \eqref{eq:bound_prob_gap_false3} is the application of the lower tail of the Chernoff bound with $\delta=(\sqrt{5}-1)/2$ for the Bernoulli distribution \cite[Theorem 4]{GoemansLectureNote}. Taking the expectation of \eqref{eq:bound_times_selected} and substitute the probability bounds for \eqref{eq:clean_event_cond1}-\eqref{eq:clean_event_cond3} to hold, we get
\begin{align}
\bbE\left[ \tilde{T}_{\ell,i^\star}[n] \right] & \le \tilde{T}_{\ell,i^\star}^{(0)} + \sum_{t=1}^{n-1} \sum_{u=1}^{t-1} \sum_{\tilde{u}_\ell=\tilde{T}_{\ell,i^\star}^{(0)}}^{t-1} (2 t^{-4} + n^{-4}) \\
& \le \tilde{T}_{\ell,i^\star}^{(0)} + 3 \sum_{t=1}^{\infty} \sum_{u=1}^{t} \sum_{\tilde{u}_\ell=1}^{t} t^{-4} \\
& = \tilde{T}_{\ell,i^\star}^{(0)} + \pi^2/2. 
\end{align}
To obtain the regret, we note that when the pair $\ell$ is selected the regret incurred is $\zeta_{\ell}\Delta_{\ell,i^\star}+(1-\zeta_{\ell})\tilde{\Delta}_{\ell,i^\star}$ because when rejected (with probability $1-\zeta_\ell$), the regret is $\tilde{\Delta}_{\ell,i^\star}$. The total expected regret is then
\begin{align}
R_2[n] \le  \sum_{\ell\in\hat{\mcB}\setminus\mcS^\star}\sum_{i^\star\in\mcS^\star}  \bbE[\tilde{T}_{\ell,i^\star}[n]] (\zeta_{\ell}\Delta_{\ell,i^\star}+(1-\zeta_{\ell})\tilde{\Delta}_{\ell,i^\star}),
\end{align}
which after rearranging terms will result in \eqref{eq:regret_bound_risk_aware_greedy_UCB}.


\section{Proof of Lemma \ref{lem:smoothness_coefficient}} \label{Apd:proof_smoothness_bound}
Assuming a normalized PAS, then the single-path PAS can be represented by the delta function $\delta(\phi-\phi^\star)$. 
The average received power can be written as
\begin{align}
\bar{\gamma}(\phi_0) = \int_{\phi_0-\Delta\phi}^{\phi_0+\Delta\phi} \delta(\phi-\phi^\star) G(\phi;\phi_0) \rmd \phi.
\end{align}
Since the gain $G(\phi;\phi_0)$ is decreasing in $[\phi_0,\phi_0+\Phi/2]$ and $|\phi^\star-\phi_0|\le\Delta\phi\le \Phi/2$ by the assumption of the Lemma, 
\begin{align}
& \frac{G(\phi;\phi_0)}{G(\phi_0+\Delta\phi;\phi_0)} \ge  1 \ge G(\phi;\phi^\star), \nonumber \\ 
& \hspace{2cm} \forall \phi\in[\phi_0-\Delta\phi,\phi_0]\cup [\phi_0,\phi_0+\Delta\phi]. 
\end{align}
Multiply both sides by $\delta(\phi-\phi^\star)$ and integrate to get 
\begin{align}
\int_{\phi_0-\Delta\phi}^{\phi_0+\Delta\phi} \delta(\phi-\phi^\star)  & G(\phi;\phi_0) / G(\phi_0+\Delta\phi;\phi_0) \rmd\phi \ge \nonumber \\  
& \int_{\phi_0-\Delta\phi}^{\phi_0+\Delta\phi}\delta(\phi-\phi^\star) G(\phi;\phi^\star)\rmd\phi \\
 \bar{\gamma}(\phi_0)/ &  G(\phi_0+\Delta\phi;\phi_0) \ge \bar{\gamma}(\phi^\star). 
\end{align}
Now, for a more general PAS $\mcP(\phi)$ with a bounded support in $[\phi_0-\Psi,\phi_0+\Psi]$, the average received power can be written~as
\begin{align}
& \bar{\gamma}(\phi_0)  = \int_{\phi_0-\Psi}^{\phi+\Psi} \mcP(\phi) G(\phi;\phi_0) \rmd \phi
\\ & = \int_{\phi_0-\Psi}^{\phi_0-\Delta\phi} \mcP(\phi) G(\phi;\phi_0) \rmd \phi
+ \int_{\phi_0-\Delta\phi}^{\phi_0+\Delta\phi} \mcP(\phi) G(\phi;\phi_0) \rmd \phi \nonumber \\
& + \int_{\phi_0+\Delta\phi}^{\phi_0+\Psi} \mcP(\phi) G(\phi;\phi_0) \rmd \phi.
\end{align} 
By the same argument as in the single-path PAS case, we have
\begin{align}
\mcP(\phi)G(\phi;\phi_0) \frac{G(\phi_0;\phi_0)}{G(\phi_0+\Delta\phi;\phi_0)} & \ge \mcP(\phi)G(\phi;\phi^\star), \nonumber \\ 
\label{eq:genPAS_ineq1}
& \hspace{-3.5cm} \forall \phi\in[\phi_0-\Delta\phi,\phi_0+\Delta\phi] \\ 
\mcP(\phi)G(\phi;\phi_0) \frac{G(\phi_0+\Psi-\Delta\phi;\phi_0)}{G(\phi_0+\Psi;\phi_0)} & \ge \mcP(\phi)G(\phi;\phi^\star), \nonumber \\ 
& \hspace{-3.5cm} \forall \phi\in[\phi_0-\Psi,\phi_0-\Delta\phi]\cup [\phi_0+\Delta\phi,\phi_0+\Psi ].  \label{eq:genPAS_ineq2}
\end{align}
Taking the integral of \eqref{eq:genPAS_ineq1} and \eqref{eq:genPAS_ineq2}, we have
\begin{align}
\bar{\gamma}(\phi_0)/G(\phi_0+\Delta\phi) \ge \bar{\gamma}(\phi^\star) - \mathsf{Err}
\end{align}
where
\begin{align}
\label{eq:Err_expression}
& \mathsf{Err} =  \left( \frac{G(\phi_0+\Psi-\Delta\phi;\phi_0)}{G(\phi_0+\Psi;\phi_0)}-\frac{1}{G(\phi_0+\Delta\phi;\phi_0)} \right) \times \nonumber \\
& \left( \int_{\phi_0-\Psi}^{\phi_0-\Delta\phi} \mcP(\phi) G(\phi;\phi_0) \rmd \phi
+ \int_{\phi_0+\Delta\phi}^{\phi_0+\Psi} \mcP(\phi) G(\phi;\phi_0) \rmd \phi \right).
\end{align}
Because $G(\phi;\phi_0)$ is decreasing and concave for $\phi\in[\phi_0,\phi_0+\Phi/2]$ by the assumptions of the Lemma (e.g., true for a uniform planar array), the coefficient is positive and decreasing as $\Delta\phi$ decreases. Since the integrands are positive by definition, $\mathsf{Err}$ is positive. 
Further, because the sum of the integrals in \eqref{eq:Err_expression} is less than $\bar{\gamma}(\phi_0)$ (thus, finite), we have $\mathsf{Err}\to 0$ as $\Delta\phi\to 0$. 
Also, for small $\Psi$ the integration intervals decrease and when $\Psi\le\Delta\phi$ they disappear, i.e., $\mathsf{Err}=0$. 


%

\ifCLASSOPTIONcaptionsoff
  \newpage
\fi



%

\bibliographystyle{IEEEtran}
\bibliography{ref}

\end{document}

%% file: new_command.tex
\usepackage{mathtools}

\newcommand*{\bH}{\boldsymbol{H}}

\newcommand*{\ba}{\boldsymbol{a}}

\newcommand*{\beff}{\boldsymbol{f}}

\newcommand*{\bh}{\boldsymbol{h}}

\newcommand*{\bw}{\boldsymbol{w}}

\newcommand*{\bone}{\boldsymbol{1}}

\newcommand*{\rmd}{\mathrm{d}}

\newcommand*{\rmm}{\mathrm{m}}

\newcommand*{\rmp}{\mathrm{p}}
\newcommand*{\rmr}{\mathrm{r}}
\newcommand*{\rms}{\mathrm{s}}
\newcommand*{\rmt}{\mathrm{t}}

\newcommand*{\rmx}{\mathrm{x}}
\newcommand*{\rmy}{\mathrm{y}}

\newcommand*{\rmT}{\mathrm{T}}

\newcommand*{\rmpl}{\mathrm{pl}}

\newcommand*{\rmopt}{\mathrm{opt}}

\newcommand*{\rmA}{\mathrm{A}}
\newcommand*{\rmD}{\mathrm{D}}

\newcommand*{\jj}{\mathrm{j}}

\newcommand{\UCB}{{\mathsf{UCB}}}

\newcommand*{\bbE}{\mathbb{E}} 
\newcommand*{\bbP}{\mathbb{P}} 

\newcommand*{\mcB}{\mathcal{B}}
\newcommand*{\mcC}{\mathcal{C}}
\newcommand*{\mcO}{\mathcal{O}}
\newcommand*{\mcP}{\mathcal{P}}
\newcommand*{\mcS}{\mathcal{S}}
\newcommand*{\mcT}{\mathcal{T}}

